\documentclass[prd,twocolumn,floatfix,amsmath,nofootinbib,amssymb,floatfix]{revtex4}
\usepackage{graphicx,color,dcolumn,booktabs,bm,multirow}
\usepackage{longtable,lscape}
\usepackage{txfonts}
\usepackage{overpic}
\usepackage{amssymb}
\usepackage{array}
\usepackage{indentfirst}
\usepackage{feynmf}   
\usepackage{slashed}  
\usepackage{cases}
\usepackage{color}
\usepackage{epstopdf}
\usepackage{booktabs}
\usepackage{tabularx}
\usepackage[colorlinks,
            citecolor=blue,
            anchorcolor=red,
            menucolor=red,
            linkcolor=red,
            filecolor=red,
            runcolor=red,
            urlcolor=blue,
            frenchlinks=red]{hyperref}

\begin{document}

\title{The light meson decays of $D$ wave charmonia}
\author{Xiao-Yu Qi$^{1}$}
\author{Xing-Dao Guo$^2$}
\author{Dian-Yong Chen$^{1,3}$\footnote{Corresponding author}}\email{chendy@seu.edu.cn}
\affiliation{
$^1$ School of Physics, Southeast University, Nanjing 210094, China\\
$^2$ \mbox{College of Physics and New Energy, Xuzhou University of Technology, Xuzhou, 221111, China}\\
$^3$ \mbox{Lanzhou Center for Theoretical Physics, Lanzhou University, Lanzhou 730000, China}\\
}
\date{\today}

\begin{abstract}
Motivated by the large non-$D\bar{D}$ branching fraction of $\psi(3770)$, we investigate the light meson decays of the $D$-wave charmonia in the present work with the meson loop mechanism. With the parameter range determined by $\psi (3770) \to \phi \eta$, our estimations of the branching fractions of the light meson decays of $\psi(3770)$ are consistent with the estimations in the literature. As for $\psi(3823)$, the partial widths of $K^\ast \bar{K}^\ast$ channel is estimated to be around 200 keV, which is the the predominant light meson decay channel of $\psi(3823)$. Our estimations also indicate that $K^{(\ast)}\bar{K}^{(\ast)}$ and $\rho \pi$ are the dominant light meson decay modes of $\psi(3842)$, while $\eta_{c2}$ dominantly decays into $\rho\rho$, $K^\ast \bar{K}^\ast$, $\omega \omega$ and $\phi\phi$. 
\end{abstract}
\pacs{****}

\maketitle

\section{introduction}

As the first established $D$ wave charmonium, $\psi(3770)$ with a mass of $(3772 \pm 6 )$ MeV and a width of $(28 \pm 5)$ MeV was first observed in the total cross sections for hadron production in $e^+e^-$ annihilation process by SLAC in 1977~\cite{Rapidis:1977cv}. The observed mass and width are in good agreement with the predictions of the $\psi(1^3D_1)$ charmonium state~\cite{Eichten:1974af, Eichten:1975ag}. Since then, this state has been observed in various process by different experimental collaborations \cite{Belle:2003eeg, BaBar:2006qlj, Belle:2007hht, BES:2007zwq, Anashin:2011kq, LHCb:2019lnr}.  Long after the observations of $\psi(3770)$, a tentative observation of a structure at a mass of 3.836 GeV in the $\pi^+ \pi^-J/\psi $ invariant mass spectrum in the $\pi Li$ scattering process was reported by the E705 Collaboration~\cite{E705:1993pry}. The first confident observation of $\psi(1^3D_2)$ was reported by the Belle Collaboration in 2013~\cite{Belle:2013ewt}. By using $772\times 10^6$ $B\bar{B}$ events collected at the $\Upsilon(4S)$ resonance with the Belle detector at the KEKB collider, the Belle Collaboration observed a new resonance in the $\chi_{c1} \gamma$ invariant mass distributions of $B\to \chi_{c1}\gamma K$ with a statistical significance of $3.8~\sigma$~\cite{Belle:2013ewt}. The observed mass was $(3823.1 \pm 1.8 \pm 0.7)$ MeV~\cite{Belle:2013ewt}, which was consistent with the theoretical expectations for the $\psi(1^3D_2)$ charmonium state~\cite{Godfrey:1985xj, Eichten:1974af, Eichten:1975ag}.  Afterwords, using proton-proton collision data collected with the LHCb detector, a new narrow charmonium state, $\psi(3842)$, was observed in the decay channels $X(3842)\to D^0D^0$ and $X(3842)\to D^+ D^{-}$~\cite{LHCb:2019lnr}. The observed mass and width were $(3842.71\pm 0.16\pm 0.12)$ MeV and $(2.79\pm 0.51\pm 0.35)$ MeV, respectively, which were consistent with the unobserved spin-3, $\psi(1^3D_3)$, charmonium state. Then spin triplet of the ground $D$ wave charmonia have been established, while the spin singlet, $\eta_{c2}(1^1D_2)$, is still unobserved. 

From the perspective of decay properties, higher charmonium should dominantly decay into a pair of charmed mesons. The observed mass of $\psi(3770)$ is about 40 MeV above the threshold of $D^0 \bar{D}^0$, and 100 MeV below $D^\ast \bar{D}$ threshold, thus, $\psi(3770)$ dominantly decays into $D\bar{D}$ final states via $P-$wave. As for $\psi(3823)$, it cannot decay into $D\bar{D}$ due to $J^P$ quantum numbers conservation in the strong decay process, while $\psi(3842)$ decays into $D\bar{D}$ via $F$ wave. Besides the open charm processes, the analysis from the BES Collaboration indicated a substantial non-$D\bar{D}$ branching fraction, which was measured to be $ (16.4\pm 7.3\pm 4.2)\%$~\cite{BES:2006fpf}, $(16.1 \pm 1.6 \pm 5.7)\%$~\cite{BES:2006dso}, and $ (15.1 \pm 5.6 \pm 1.8)\%$~\cite{BES:2008vad}, with variations attributed to different analysis techniques and independent data samples. In addition to $\psi(3770)$, the $\psi(3823)$ cannot decay into $D\bar{D}$, and $\psi(3842)$ couples to $D\bar{D}$ only via $P$ wave, thus, the non-$D\bar{D}$ decay processes are even more important for $\psi(3823)$ and $\psi(3842)$. Moreover, the unobserved $\eta_{c2}$ cannot decay into $D\bar{D}$ due to the $J^P$ quantum numbers conservation. Its mass should be below the threshold of $J/\psi \omega$, which indicates that the hidden charm decay processes should also be forbidden for $\eta_{c2}$. Thus the investigations of the light meson decays of $\eta_{c2}$ are particularly interesting, which may provide important information for the observation of $\eta_{c2}$.

For the light meson decay process of charmonia, the charmed and anti-charmed quark should annihilate into gluons and then the gluons create two light quark pairs, and these quark pairs transit into light mesons in the final states, such kind process should be suppressed due to Okubo-Zweig-Iizuka (OZI) rule. However, some experimental measurements indicate that the branching fractions of such kind of OZI suppressed processes are much greater than expectations. To understand this phenomenon, the meson loop mechanism, which is considered as the phenomenological description of the long range interactions, had been introduced in various processes~\cite{Chen:2009ah, Li:2013zcr, Zhang:2009kr, Chen:2014sra, Chen:2014ccr, Liu:2009dr, Qi:2025hwd}. In our previous work, the hidden charm decay processes, as one kind of important non-$D\bar{D}$, have been investigated by using the meson loop mechanism~\cite{Qi:2025hwd}. In the present work, we further extend the meson loop mechanism to investigate the light meson decays of the $D$-wave charmonia, such as $\psi(3770)/\psi(3823)/\psi(3842) \to \mathcal{VV,\ VP,\ PP}$ and $\eta_{c2}(1D) \to \mathcal{VV, PV}$.

This work is organized as follows. After the introduction, the meson loop mechanism is introduced and the amplitudes of relevant processes are obtained by using an effective Lagrangian approach in Section~\ref{sec:Sec2}. In Section ~\ref{Sec:Num}, we present the numerical results of the light meson decays of the $D$-wave charmonia and the relevant discussions, while the last section is devoted to a short summary.

\section{The light meson decays of the $D-$wave charmonia}
\label{sec:Sec2}
\subsection{Effective Lagrangians}
As discussed above, the meson loop mechanism is crucial to understand the OZI suppressed decays of the charmonia. In the present work, the meson loops are estimated in the hadron level, and the interaction between the involved hadrons are depicted by effective Lagrangians, which are constructed by the heavy quark limit and chiral symmetry. In the heavy limit, the heavy-light meson can be characterized by the freedom of light degrees, i.e., $\vec{s}_{\ell}=\vec{s}_q+\vec{\ell}$, where $\vec{s}_q$ and $\vec{\ell}$ are the light quark spin and the relative orbital angular momentum, respectively. In this limit, the wave function of the heavy-light mesons are independent on the spin and flavor of the heavy quark, and the meson with the same $s_\ell$ are degenerated. For example, for the $S-$wave heavy-light meson with $\vec{s}_\ell=1/2$ can be expressed in the matrix form, which is~\cite{Lipkin:1986av,Casalbuoni:1996pg,Kaymakcalan:1983qq,Oh:2000qr}, 
\begin{eqnarray}
H_1&=&\frac{1+v\!\!\!\slash}{2} \Big[ \mathcal{D}^{\ast\mu}\gamma_\mu- \mathcal{D}\gamma_5 \Big] ,\nonumber\\
H_2&=&\Big[\bar{\mathcal{D}}^{\ast\mu}\gamma_\mu-\bar{\mathcal{D}}\gamma_5\Big]\frac{1+v\!\!\!\slash}{2},\nonumber\\
\bar{H}_{1,2}&=&\gamma^0 H^\dagger_{1,2}\gamma^0,
\end{eqnarray}
respectively, with $\mathcal{D}^{(\ast)}=\left(D^{(\ast)0} D^{(\ast) +}, D_s^{(\ast) +}\right)$.

For the heavy quarkonium, the heavy quark flavor symmetry does not hold any more since the kinetic energy operator includes a factor of $1/m_Q$, but the degeneracy is still expected under the rotations of the two heavy quark spin, thus one can build up heavy quarkonium multiplets for each value of relative angular momentum $\ell$. For the $D$-wave charmonia, the multiplets can be expressed as~\cite{Lipkin:1986av},
\begin{eqnarray}
\mathcal{J}^{\mu\lambda}&=&\frac{1+v\!\!\!\slash}{2}\Big[\psi^{\mu\alpha\lambda}_3 \gamma_\alpha+\frac{1}{\sqrt{6}}\big(\psi^{\mu\alpha\beta\rho}v_\alpha \gamma_\beta \psi^\lambda_{2\rho}+\psi^{\lambda\alpha\beta\rho}v_\alpha \gamma_\beta \psi^\mu_{2\rho}\big)\nonumber\\
&&+\frac{\sqrt{15}}{10}\big[(\gamma^\mu-v^\mu)\psi^\lambda_1+(\gamma^\lambda-v^\lambda)\psi^\mu_1 \big ]\nonumber\\
&&-\frac{1}{\sqrt{15}}(g^{\mu\lambda}-v^\mu v^\lambda)\gamma_\alpha \psi^\alpha_1+\eta^{\mu\lambda}_{c2}\gamma_5\Big]\frac{1-v\!\!\!\slash}{2}.
\end{eqnarray}

The effective interactions between the $D-$wave charmonia and the $S-$wave charmed meson pair can be constructed as,
\begin{eqnarray}
\mathcal{L}=ig_1 \mathrm{Tr}\Big[\mathcal{J}^{\mu\lambda}\bar{H}_2 {\stackrel{\leftrightarrow}{\partial_\mu}}\gamma_\lambda \bar{H}_1\Big]+\mathrm{H.c.}.
\end{eqnarray}

With the above effective Lagrangian, we can obtain the specific form of the effective interactions relevant to the current calculations, denoted as,
\begin{eqnarray}
\mathcal{L}&=&g_{\psi_1 \mathcal{D} \mathcal{D} }\psi^\mu_1 \big( \mathcal{D} \partial_\mu \mathcal{D}^\dagger-\mathcal{D}^\dagger\partial_\mu \mathcal{D}\big) \nonumber\\
&&+g_{\psi_1 \mathcal{D} \mathcal{D}^\ast}\varepsilon_{\mu\nu\alpha\beta}\big[ \mathcal{D}{\stackrel{\leftrightarrow}{\partial^\mu}} \mathcal{D}^{\ast\beta\dagger}- \mathcal{D}^{\ast\beta}{\stackrel{\leftrightarrow}{\partial^\mu}} \mathcal{D}^\dagger\big]\partial^\nu \psi_{1}^\alpha\nonumber\\
&&+g_{\psi_1 \mathcal{D}^\ast \mathcal{D}^\ast}\big[-4(\psi^\mu_1 \mathcal{D}^{\ast\nu\dagger}\partial_\mu \mathcal{D}^{\ast}_\nu-\psi^\mu_1 \mathcal{D}^{\ast}_\nu \partial_\mu \mathcal{D}^{\ast\nu\dagger})\nonumber\\
&&\hspace{1.5cm}+\psi^\mu_1 \mathcal{D}^{\ast\nu\dagger}\partial_\nu \mathcal{D}^{\ast}_\mu-\psi^\mu_1 \mathcal{D}^{\ast\nu}\partial_\nu \mathcal{D}^{\ast\dagger}_\mu\big]\nonumber\\
&&+ig_{\psi_2 \mathcal{D} \mathcal{D}^\ast}\psi^{\mu\nu}_2 \big(\mathcal{D}{\stackrel{\leftrightarrow}{\partial_\nu}} \mathcal{D}^{\ast\dagger}_\mu- \mathcal{D}^{\ast}_\mu{\stackrel{\leftrightarrow}{\partial_\nu}} \mathcal{D}^\dagger \big)\nonumber\\
&&+ig_{\psi_2 \mathcal{D}^\ast \mathcal{D}^\ast}\varepsilon_{\alpha\beta\mu\nu} \big[ \mathcal{D}^{\ast\nu}{\stackrel{\leftrightarrow}{\partial^\beta}} \mathcal{D}^{\ast\dagger}_\lambda- \mathcal{D}^{\ast\nu\dagger}{\stackrel{\leftrightarrow}{\partial^\beta}} \mathcal{D}^{\ast}_\lambda \big]\partial^\mu \psi^{\alpha\lambda}_2\nonumber\\
&&+g_{\psi_3 \mathcal{D}^\ast \mathcal{D}^\ast}\psi^{\mu\nu\alpha}_3 \big[ \mathcal{D}^{\ast}_\alpha{\stackrel{\leftrightarrow}{\partial_\mu}}\mathcal{D}^{\ast\dagger}_\nu +\mathcal{D}^{\ast}_\nu{\stackrel{\leftrightarrow}{\partial_\mu}}\mathcal{D}^{\ast\dagger}_\alpha \big]\nonumber\\
&&+g_{\eta_{c2} \mathcal{D}^*\mathcal{D}}\eta_{c2}^{\mu\alpha} \big(\mathcal{D}^*_\alpha\stackrel{\leftrightarrow}{\partial_\mu}\mathcal{D}^{\dagger}+\mathcal{D}\stackrel{\leftrightarrow}{\partial_\mu}\mathcal{D}^{*\dagger}_\alpha\big)\nonumber\\
&&+ig_{\eta_{c2} \mathcal{D}^*\mathcal{D}^*}\varepsilon_{\alpha\beta\lambda\rho}\mathcal{D}^{*\alpha}\stackrel{\leftrightarrow}{\partial_\mu}\mathcal{D}^{*\beta\dagger}\partial^{\rho}\eta_{c2}^{\mu\lambda}.
\end{eqnarray}

Taking into account the chiral symmetry and the heavy quark limit, the effective interactions about pseudoscalar mesons and vector mesons with charm mesons are constructed as~\cite{Falk:1992cx, Chen:2014sra, Yan:1992gz, Cheng:1992xi, Wise:1992hn},
\begin{eqnarray}
\mathcal{L}_{\mathcal{D}^{(*)} \mathcal{D}^{(*)}\mathcal{P}}&=&-ig_{\mathcal{D}^\ast \mathcal{D} \mathcal{P}}\big(\mathcal{D}^{i\dagger} \partial^\mu \mathcal{P}_{ij}\mathcal{D}^{\ast j}_\mu-\mathcal{D}^{\ast i\dagger}_\mu \partial^\mu \mathcal{P}_{ij} \mathcal{D}^{j}\big)\nonumber\\
&&+\frac{1}{2}g_{\mathcal{D}^\ast \mathcal{D}^\ast \mathcal{P}}\varepsilon_{\mu\nu\alpha\beta} \mathcal{D}^{\ast \mu\dagger}_i \partial^\nu \mathcal{P}_{ij}{\stackrel{\leftrightarrow}{\partial^\alpha}}\mathcal{D}^{\ast \beta}_j,\nonumber\\
\mathcal{L}_{\mathcal{D}^{(*)} \mathcal{D}^{(*)}\mathcal{V}}&=&-i g_{\mathcal{D} \mathcal{D} \mathcal{V}} \mathcal{D}_{i}^{\dagger} \stackrel{\leftrightarrow}\partial_{\mu} \mathcal{D}^{j}\big(\mathcal{V}^{\mu}\big)_{j}^{i}-2 f_{\mathcal{D}^{*} \mathcal{D} \mathcal{V}} \varepsilon^{\mu \nu \alpha \beta}\nonumber \\
&& \times \big(\partial_{\mu} \mathcal{V}_{\nu}\big)_{j}^{i}\big(\mathcal{D}_{i}^{\dagger}\stackrel{\leftrightarrow} \partial_{\alpha} \mathcal{D}_{\beta}^{* j}-\mathcal{D}_{\beta i}^{* \dagger} \stackrel{\leftrightarrow}{\partial}_{\alpha} \mathcal{D}^{j}\big)\nonumber \\
&&+i g_{\mathcal{D}^{*} \mathcal{D}^{*} \mathcal{V}} \mathcal{D}_{i}^{* \dagger \nu} \stackrel{\leftrightarrow}{\partial}_{\mu} \mathcal{D}_{\nu}^{* j}\big(\mathcal{V}^{\mu}\big)_{j}^{i}\nonumber \\
&&+4 i f_{\mathcal{D}^{*} \mathcal{D}^{*} \mathcal{V}} \mathcal{D}_{i}^{* \dagger \mu}\big(\partial_{\mu} \mathcal{V}^{\nu}-\partial^{\nu} \mathcal{V}_{\mu}\big)_{j}^{i} \mathcal{D}_{\nu}^{* j}.
\end{eqnarray}
where $\mathcal{P}$ and $\mathcal{V}$ are the matrix form of the light pseudoscalar and vector mesons, and their concrete forms are~\cite{Casalbuoni:1996pg,Casalbuoni:1992gi,Casalbuoni:1992dx}, 
\begin{eqnarray}
\mathcal{P} &=& \left(\begin{array}{ccc}
\frac{\pi^{0}}{\sqrt{2}}+\alpha \eta+\beta \eta \prime & \pi^{+} & K^{+} \\
\pi^{-} & -\frac{\pi^{0}}{\sqrt{2}}+\alpha \eta+\beta \eta \prime & K^{0} \\
K^{-} & \bar{K}^{0} & \gamma \eta+\delta \eta^{\prime}
\end{array}\right), \nonumber\\
\nonumber\\ \nonumber\\
\mathcal{V} &=& \left(\begin{array}{ccc}
\frac{1}{\sqrt{2}}\left(\rho^{0}+\omega\right) & \rho^{+} & K^{*+} \\
\rho^{-} & \frac{1}{\sqrt{2}}\left(-\rho^{0}+\omega\right) & K^{* 0} \\
K^{*-} & \bar{K}^{* 0} & \phi
\end{array}\right),
\end{eqnarray}
where the parameters $\alpha$, $\beta$, $\gamma$ and $\delta$ can relate to the mixing angle $\theta$ by~\cite{Gilman:1987ax}, 
\begin{eqnarray}
\alpha &=& \frac{\cos \theta-\sqrt{2} \sin \theta}{\sqrt{6}}, \quad \beta=\frac{\sin \theta+\sqrt{2} \cos \theta}{\sqrt{6}} ,\nonumber\\
\gamma &=& \frac{-2 \cos \theta-\sqrt{2} \sin \theta}{\sqrt{6}}, \quad \delta=\frac{-2 \sin \theta+\sqrt{2} \cos \theta}{\sqrt{6}},
\end{eqnarray}
with the mixing angle to be $\theta = 19.1^{\circ}$~\cite{MARK-III:1988crp, DM2:1988bfq}.

\subsection{Amplitudes for $\psi(3770)\to \mathcal{PV}$}

\begin{figure*}[htb]
	\begin{tabular}{cccccc}
		\centering
		\includegraphics[width=2.8cm]{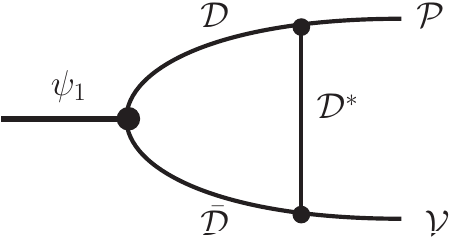}&
		\includegraphics[width=2.8cm]{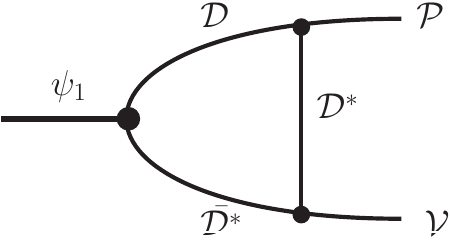}&
		\includegraphics[width=2.8cm]{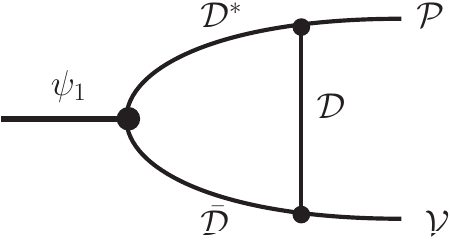}&
		\includegraphics[width=2.8cm]{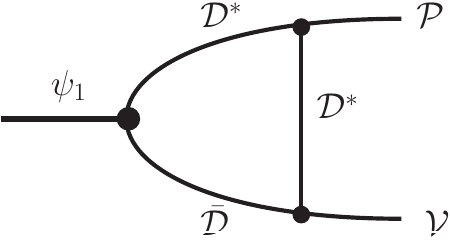}&
		\includegraphics[width=2.8cm]{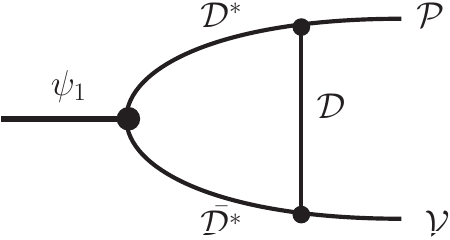}&
		\includegraphics[width=2.8cm]{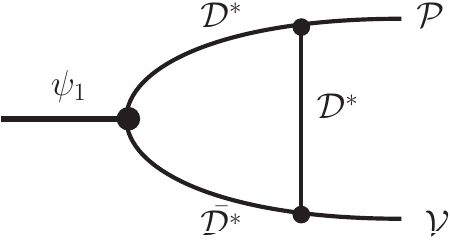}\\
		\\
		$(A_1)$ & $(A_2)$ & $(A_3)$ & $(A_4)$ & $(A_5)$ & $(A_6)$ \\
	\end{tabular}
	\caption{Diagrams contributing to $\psi(3770)\to \mathcal{PV}$ at the hadron level.\label{Fig:Tri11}}
\end{figure*}

Here we take $\psi(3770) \to \mathcal{P}\mathcal{V}$ as examples, $\psi(3770)$ could couples to $\mathcal{D}^{(\ast)} \bar{\mathcal{D}}^{(\ast)}$, and this charmed meson pair transits into light pseudoscalar and vector mesons by exchanging a proper charmed meson. The sketch diagrams contributing to $\psi(3770) \to \mathcal{PV}$ are listed in Fig.~\ref{Fig:Tri11} and the corresponding amplitudes constructed by effective Lagrangians read,
\begin{widetext}
\begin{eqnarray}
\mathcal{M}_{A_1}&=&i^3 \int\frac{d^4 q}{(2\pi)^4}\left[
	ig_{\psi_{1} \mathcal{D} \mathcal{D} }\epsilon^{\mu}(p)\left(p_{2 \mu}- p_{1 \mu}\right)\right] \left[-g_{ \mathcal{D}^{*} \mathcal{D}P} p_{3\chi }\right]\left[2f_{\mathcal{D}^*\mathcal{D}V}\varepsilon_{\kappa\tau\phi\xi}p_4^\kappa\epsilon^\tau(p_4)\big(p_2^\phi-q^\phi\big)\right]
	\nonumber\\
	&&\times\frac{1}{p^2_1-m^2_1}\frac{1}{p^2_2-m^2_2}\frac{-g^{\chi\xi}+q^{\chi} q^{\xi} /m^2_q}{q^2-m^2_q}\mathcal{F}^2(m_{q}^2, q^2)\nonumber\\
\mathcal{M}_{A_2}&=&i^3 \int\frac{d^4 q}{(2\pi)^4}\left[
	g_{\psi_1 \mathcal{D}^{*} \mathcal{D}} \varepsilon_{\mu \nu \alpha \beta}\left(p^{\mu}_{2}-p^{\mu}_{1}\right)  p^{v}\epsilon^{\alpha}\left(p\right)\right] \left[-g_{\mathcal{D}^{*} \mathcal{D}P} p_{3\chi }\right]\left[g_{\mathcal{D}^{*} \mathcal{D}^{*}V}g_{\tau \phi} \epsilon^{\tau}\left(p_{4}\right)\left(q^{\xi}-p^{\xi}_{2}\right)\right.\nonumber\\
	&&\left.-4f_{\mathcal{D}^{*} \mathcal{D}^{*}V} \epsilon^{\tau}\left(p_{4}\right)\left(p_4^{\xi}g_{\tau \phi}-p^{\phi}_{4}g_{\tau \xi}\right)\right]\frac{1}{p^2_1-m^2_1} \times\frac{-g^{\beta\xi}+p_2^{\beta} p_2^{\xi} /m^2_2}{p^2_2-m^2_2}\frac{-g^{\chi\phi}+q^{\chi} q^{\phi} /m^2_q}{q^2-m^2_q}\mathcal{F}^2(m_{q}^2 , q^2)\nonumber\\
\mathcal{M}_{A_3}&=&i^3 \int\frac{d^4 q}{(2\pi)^4}\left[
	g_{\psi_{1} \mathcal{D}^{*} \mathcal{D}} \varepsilon_{\mu \nu \alpha \beta} \left(p^{\mu}_{1}-p^{\mu}_{2}\right)  p^{\nu} \epsilon^{\alpha}(p)\right] \left[g_{\mathcal{D}^{*} \mathcal{D}P} p_{3\chi }\right]\left[g_{ \mathcal{D}\mathcal{D}V} \epsilon^{\tau}\left(p_{4}\right) \left(p_{2\tau}-q_{\tau}\right)\right]
	\nonumber\\
	&&\times\frac{-g^{\beta\chi}+p_1^{\beta} p_1^{\chi} /m^2_1}{p^2_1-m^2_1}\frac{1}{p^2_2-m^2_2}\frac{1}{q^2-m^2_q}\mathcal{F}^2(m_{q}^2, q^2)\nonumber\\
\mathcal{M}_{A_4}&=&i^3 \int\frac{d^4 q}{(2\pi)^4}\left[
	g_{\psi_{1} \mathcal{D}^{*} \mathcal{D}} \varepsilon_{\mu \nu \alpha \beta} \left(p^{\mu}_{1}-p^{\mu}_{2}\right)  p^{\nu} \epsilon^{\alpha}(p)\right] \left[\frac{1}{2} g_{ \mathcal{D}^{*} \mathcal{D}^{*}P } \varepsilon_{\chi\lambda\sigma\rho}  p_{3}^{\lambda} \left(p^{\sigma}_{1}+q^{\sigma}\right) \right]\left[2f_{\mathcal{D}^*\mathcal{D}V}\varepsilon_{\kappa\tau\phi\xi} p_4^\kappa\epsilon^\tau(p_4)\big(p_2^\phi-q^\phi\big)\right] \nonumber\\
	&&\frac{1}{p^2_2-m^2_2} \frac{-g^{\beta\rho}+p_1^{\beta} p_1^{\rho} /m^2_1}{p^2_1-m^2_1}\frac{-g^{\chi\xi}+q^{\chi} q^{\xi} /m^2_q}{q^2-m^2_q}\mathcal{F}^2(m_{q}^2, q^2)\nonumber\\
\mathcal{M}_{A_5}&=&i^3 \int\frac{d^4 q}{(2\pi)^4}\left[ig_{\psi_{1} \mathcal{D}^{*} \mathcal{D}^{*}}\epsilon^{\mu}(p)\left(-4 ( p_{1\mu}-p_{2\mu})  g^{\alpha \beta} +p^{\alpha}_{1}g_{\mu}^{\beta}-p^{\beta}_{2}g^{\alpha}_{\mu}\right) \right] \left[g_{\mathcal{D}^{*} \mathcal{D}P}p_{3\chi }\right] \left[2f_{\mathcal{D}^*\mathcal{D}V}\varepsilon_{\kappa\tau\phi\xi}p_4^\kappa\epsilon^\tau(p_4)\big(q^\phi-p_2^\phi\big)\right] \nonumber\\
	&&\times\frac{-g^{\beta\chi}+p_1^{\beta} p_1^{\chi} /m^2_1}{p^2_1-m^2_1}\frac{-g^{\alpha\xi}+p_2^{\alpha} p_2^{\xi} /m^2_2}{p^2_2-m^2_2} \frac{1}{q^2-m^2_q}\mathcal{F}^2(m_{q}^2, q^2)\nonumber\\
\mathcal{M}_{A_6}&=&i^3 \int\frac{d^4 q}{(2\pi)^4}\left[ig_{\psi_{1} \mathcal{D}^{*} \mathcal{D}^{*}}\epsilon^{\mu}(p)\left(-4 ( p_{1\mu}-p_{2\mu})  g^{\alpha \beta}+p^{\alpha}_{1}g_{\mu}^{\beta}-p^{\beta}_{2}g^{\alpha}_{\mu}\right) \right]  \left[\frac{1}{2} g_{\mathcal{D}^{*} \mathcal{D}^{*}P } \varepsilon_{\chi\lambda\sigma\rho}  p_{3}^{\lambda} (p^{\sigma}_{1}+q^{\sigma}) \right]\Big[g_{\mathcal{D}^{*} \mathcal{D}^{*}V}g_{\tau \phi} \epsilon^{\tau}\left(p_{4}\right)\left(q^{\xi}-p^{\xi}_{2}\right) 
	\nonumber\\
	&&-4f_{\mathcal{D}^{*} \mathcal{D}^{*}V}\epsilon^{\tau}\left(p_{4}\right)\left(p_4^{\xi}g_{\tau \phi}-p^{\phi}_{4}g_{\tau \xi}\right)\Big]
\times\frac{-g^{\beta\rho}+p_1^{\beta} p_1^{\rho} /m^2_1}{p^2_1-m^2_1}\frac{-g^{\alpha\xi}+p_2^{\alpha} p_2^{\xi} /m^2_2}{p^2_2-m^2_2}\frac{-g^{\chi\phi}+q^{\chi} q^{\phi} /m^2_q}{q^2-m^2_q}\mathcal{F}^2(m_{q}^2, q^2).
\end{eqnarray}
\end{widetext}
It should be noted that in the amplitudes mentioned above, a form factor ${F}(m^2,q^2)$ is introduced to ensure the convergence of the amplitude in the ultraviolet region and to account for off-shell effects. In the present work, a form factor in the monopole form is employed, which is,
\begin{eqnarray}
\mathcal{F}\left(m, q^{2}\right)=\left(\frac{m^{2}-\Lambda^{2}}{q^{2}-\Lambda^{2}}\right),
\end{eqnarray}
where the parameter $\Lambda$ can be reparameterized as $\Lambda=m_{E}+\alpha \Lambda_{\mathrm{QCD}}$ with $m_E$ to be the mass of the exchanged meson and $\Lambda_{\mathrm{QCD}}=220$ MeV. In general, the model parameter $\alpha$ should be uniform, however its precise value cannot be derived using first principle~\cite{Tornqvist:1993vu,Tornqvist:1993ng,Locher:1993cc,Li:1996yn}. When dealing with specific processes, a more common practice is to determine the exact value of $\alpha$ by comparing experimental measurements with theoretical estimations.

The total amplitude for $\psi(3770)\to \mathcal{PV}$ is,
\begin{eqnarray}
    \mathcal{M}^{\mathrm{Tot}}_{\psi_1 \to \mathcal{PV}} =\sum_{i=1}^{5} \mathcal{M}_{A_i}.
\end{eqnarray}
Then, the paritial width of $\psi(3770)\to\mathcal{P}\mathcal{V}$ can be estimated by,
\begin{eqnarray}
&&\Gamma(\psi_1 \to \mathcal{P}\mathcal{V})=\frac{1}{3} \frac{1}{8 \pi} \frac{|\vec{p}|}{m_{0}^{2}}\overline{\left|\mathcal{M}_{\psi_{1} \to\mathcal{P}\mathcal{V}}^{\mathrm{tot}}\right|^{2}},\nonumber\\
\end{eqnarray}
where the factor $1/3$ comes from average of the spin of the initial state and $|\vec{p}|$ denotes the value of the three-momentum of the final state in the initial rest frame. 

In the same way, we can construct the amplitdues for $\psi(3770) \to \mathcal{VV}/\mathcal{PP}$ , $\psi(3823) \to \mathcal{PV}/\mathcal{VV}/\mathcal{PP}$ , $\psi(3842) \to \mathcal{PV}/\mathcal{VV}/\mathcal{PP}$ and $\eta_{c2} \to \mathcal{VV}/\mathcal{PV}$, which are collected in Appendix~\ref{Sec:Apppsi1}-\ref{Sec:Appetac2}.

\section{Numerical Results and discussions}
\label{Sec:Num}

\subsection{Coupling constants}
Before estimating the partial widths of the involved processes, the relevant coupling constants should be further clarified. In the heavy limit, the coupling constants related to the $D-$wave charmonia and charmed meson pairs can relate to the gauge coupling constants $g_1$ by, 
\begin{eqnarray}
g_{\psi_1 \mathcal{D}\mathcal{D}}&=&-2g_1 \frac{\sqrt{15}}{3}\sqrt{m_{\psi_1}m_\mathcal{D} m_\mathcal{D}},\nonumber\\
g_{\psi_1 \mathcal{D}\mathcal{D}^\ast}&=&g_1 \frac{\sqrt{15}}{3}\sqrt{m_{\mathcal{D}}m_{\mathcal{D}^\ast}/m_{\psi_1}},\nonumber\\
g_{\psi_1 \mathcal{D}^\ast \mathcal{D}^\ast}&=&-g_1 \frac{\sqrt{15}}{15}\sqrt{m_{\psi_1}m_{\mathcal{D}^\ast} m_{\mathcal{D}^\ast}},\nonumber\\
g_{\psi_2 \mathcal{D}\mathcal{D}^\ast}&=&2g_1 \sqrt{\frac{3}{2}}\sqrt{m_{\psi_2}m_\mathcal{D} m_{\mathcal{D}^\ast}},\nonumber\\
g_{\psi_2 \mathcal{D}^\ast \mathcal{D}^\ast}&=&-2g_1 \sqrt{\frac{1}{6}}\sqrt{m_{\mathcal{D}^\ast}m_{\mathcal{D}^\ast}/m_{\psi_2}},\nonumber\\
g_{\psi_3 \mathcal{D}^\ast \mathcal{D}^\ast}&=&2g_1 \sqrt{m_{\psi_3}m_{\mathcal{D}^\ast} m_{\mathcal{D}^\ast}},\nonumber\\
g_{\eta_{c2} \mathcal{D}^{*} \mathcal{D}}&=&2 g_{1} \sqrt{m_{\eta_{c2}} m_{\mathcal{D}} m_{\mathcal{D}^{*}}} \nonumber\\
g_{\eta_{c2} \mathcal{D}^{*} \mathcal{D}^{*}}&=&2 g_{1} m_{\mathcal{D}^{*}} / \sqrt{m_{\eta_{c2}}},
\end{eqnarray}
with the gauge coupling $g_1 = 1.39$, which is determined by the partial width of $\psi(3770)\to D\bar{D}$ ~\cite{BES:2008vad}.

Considering the heavy quark limit and chiral symmetry, one can obtain the coupling constants relevant to the light pseudoscalar and vector mesons, which are~\cite{Falk:1992cx,Chen:2014sra,Yan:1992gz,Cheng:1992xi,Wise:1992hn},
\begin{eqnarray}
g_{\mathcal{D}^{\ast}\mathcal{D}P}&=&\frac{2g}{f_\pi}\sqrt{m_{\mathcal{D}^\ast}m_{\mathcal{D}}},\nonumber\\
g_{\mathcal{D}^{\ast}\mathcal{D}^\ast P}&=&\frac{2g}{f_\pi},\nonumber\\
g_{\mathcal{D}\mathcal{D}V}&=&g_{\mathcal{D}^*\mathcal{D}^*V}=\frac{1}{\sqrt2}\beta g_v,\nonumber\\
f_{\mathcal{D}^*\mathcal{D}V}&=&\frac{f_{\mathcal{D}^*\mathcal{D}^*V}}{m_\mathcal{D}^*}=\frac{1}{\sqrt2}\lambda g_v.
\end{eqnarray}
where $\beta_V=0.9$, $\lambda_V=0.56$, and $g_V={m_\rho}/{f_\pi}$ with $f_\pi=132\  \mathrm{MeV}$ denoting to the decay constant of pion, and $g=0.59$, which is evaluated by the partial width of $D^\ast\to D\pi$~\cite{ParticleDataGroup:2024cfk}.  

\begin{figure*}[t]
	\centering
	\begin{tabular}{ccc}
\includegraphics[width=8.4cm]{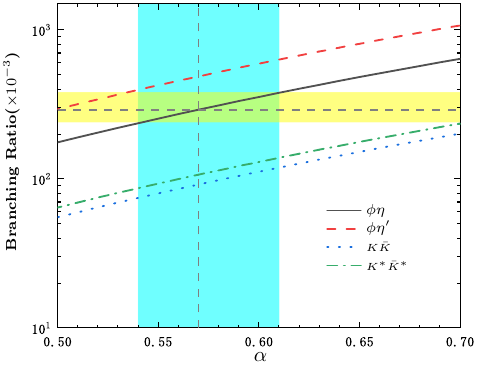} & \hspace{0.5cm} &
\includegraphics[width=8.4cm]{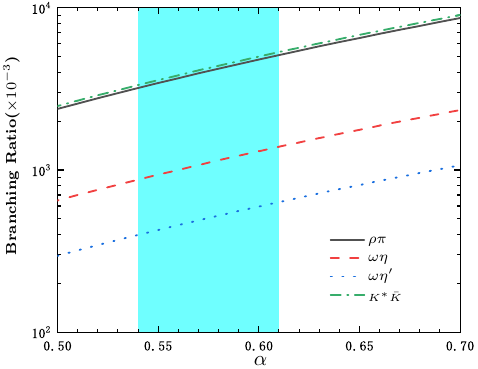}
	\end{tabular}
	\caption{The branching fractions of the light meson decays of $\psi(3770) \to $ depending on the model parameter $\alpha$. The horizontal gray dashed line with yellow  band is the measured branching fraction of $\psi(3770) \to \phi\eta$, while the vertical gray dashed line with cyan band indicates the $\alpha$ range determined by comparing our estimations with the experimental measurement.}\label{Fig:Br1}
\end{figure*}

\begin{table*}[htb]
\renewcommand\arraystretch{1.5}
\centering
\caption{the branching fractions of light meson decays of $\psi(3770)  $. For comparison, the results from other references~\cite{Li:2013zcr, Zhang:2009kr} and the experimental measurements~\cite{ParticleDataGroup:2024cfk} are also listed. In addition, the branching fractions of $K^{\ast+} K^- +cc.$ and $K^{\ast 0} \bar{K}^0$ are listed separately for comparison.} \label{t1}
\begin{tabular}{p{3cm}<\centering p{2.5cm}<\centering p{2.5cm}<\centering p{2.5cm}<\centering p{2.5cm}<\centering p{2.5cm}<\centering}
\toprule[1pt]
BR($\times 10^{-4}$)   &  Present    &Ref.~\cite{Li:2013zcr}      &Ref.~\cite{Zhang:2009kr}   &PDG~\cite{ParticleDataGroup:2024cfk}  &Ratio \\
\midrule[1pt]
$\phi\eta$               &2.40-3.80 (Input)      &0.14-9.28      &1.25              &2.4-3.8           & 1.0 \\
$\rho \pi$  &32.10-51.00      &2.16-168        &34.44             &$<$0.048          & 13.5 \\
$\omega\eta$             &8.71-13.9      &0.50-39.9       &6.83              &$<$0.14           & 3.7\\
$\omega\eta^\prime$      &3.97-6.33      &0.22-16.7      &0.58              &$<$4              & 1.7\\
$\phi\eta^\prime$        &3.95-6.32      &0.33-21.6      &0.87              &$<$0.23           & 1.7 \\
$K^{*+}K^-+c.c.$          &16.68-26.64    &0.96-84.4      &10.97             &$<$0.14           & 7.1 \\
$K^{*0}\bar{K^0}+c.c.$    &16.47-26.32    &0.78-85.0      &11.8              &$<$12             & 7.0 \\
$K\bar{K}$               &0.75-1.19      &-              &-                 &-                 & 0.3 \\
$K^{*}\bar{K^*}$         &0.87-1.39      &-              &-                 &-                 & 0.4 \\
\bottomrule[1pt]
\end{tabular}
\end{table*}

\subsection{light meson decays of $\psi(3770)$}

Besides the above coupling constants, only one parameter, $\alpha$, introduced by the form factor, is undetermined. It's worth mentioning that the process $\psi(3770) \to \phi \eta$ has been measured by the BES and CLEO Collaborations~\cite{CLEO:2005zrs, BES:2007zan}, and the PDG average of the branching fraction is $(3.1\pm 0.7)\times 10^{-4}$~\cite{ParticleDataGroup:2024cfk}. With the branching fraction of $\psi(3770) \to \phi \eta$, we can determine the model parameter range by comparing the theoretical estimation in the present work with the experimental data, and within this parameter range, the branching fractions or the partial widths of the discussed process could be evaluated.

Our estimations of the branching fractions of the light meson decays of $\psi(3770)$ depending on the model parameter $\alpha$ are presented in Fig.~\ref{Fig:Br1}. The horizontal gray dashed line with yellow  band is the measured branching fraction of $\psi(3770) \to \phi\eta$, with this branching fraction, we can determine the model parameter range, which is  $0.54< \alpha <0.61$, shown as the vertical gray dashed line with cyan band. From the figure, one can find that the branching fractions of most channels are greater than the that of $\psi(3770) \to \phi \eta$ except for the  $K\bar{K}$ and $K^\ast \bar{K}^\ast$ channels. Among these processes, the branching fraction of $psi(3770) \to K^\ast \bar{K}^{\ast}$ is the largest one, and the branching ratio for $\rho \pi^0$ and $\omega \eta$ are similar, which is about half of that of $psi(3770) \to K^\ast \bar{K}^{\ast}$. In addition, our estimations show that the $\alpha$ dependences of the branching fractions of the involved process are very similar, thus, their ratios are expected to be weakly dependent on the model parameter $\alpha$.

 With the model parameter range determined by $\psi(3770) \to \phi \eta$, one can estimate the branching fractions of other light meson decay channels of $\psi(3770)$, which are collected in Table~\ref{t1}. Here we take the process of $\psi(3770) \to \phi \eta$ as scale, the ratios of the branching fractions of other processes to that of $\psi(3770)\to \phi \eta$ are listed in the last column of Table~\ref{t1}. Our estimations indicate that the branching fractions of $\rho \pi$ (including $\rho^+ \pi^-$, $\rho^0 \pi^0$ and $\rho^- \pi^+$) and $K^\ast \bar{K}+c.c.$ (including $K^{\ast+} K^{-}$, $K^{\ast 0} \bar{K}^0$, $K^+ K^{\ast -}$ and $K^{0} \bar{K}^{\ast 0}$ ) are about 13.5 and $14.1$ times of that $\phi \eta$ channel, respectively. The branching fractions of $\omega \eta$, $\omega \eta^\prime$ and $\phi \eta^\prime$ are the same order as the one of $\phi \eta$ channel, while the branching fractions of $K\bar{K}$ and $K^\ast \bar{K}^\ast $ are several times smaller than that of $\phi \eta$.

 For comparison, the estimations from other group are also collected in Table~\ref{t1}~\cite{Li:2013zcr, Zhang:2009kr}. It should be noted that in both Refs.~\cite{Li:2013zcr, Zhang:2009kr} and the present work, the meson-loop contributions have been considered. In Ref.~\cite{Li:2013zcr}, the model parameter $\alpha$ was taken in a very large range, then the obtained branching fractions were investigated with a very large uncertainties, and the branching fractions in the present estimations with a narrow model parameter are all in the range of those in Ref.~\cite{Li:2013zcr}. As for Ref.~\cite{Zhang:2009kr}, the authors considered the $t$ and $s$-channel meson loops, and the pQCD leading transition via short-range gluon exchanges as well. By reproducing the center value of the branching fraction of $\psi(3770) \to \phi \eta$ with the total contributions, they determined the model parameter and the branching fractions listed in Table ~\ref{t1} are the results from the meson loop contributions, which could be compared with our results directly. Comparing to Ref.~\cite{Zhang:2009kr}, the present estimations have considered more diagrams, such as diagram $A_5$ and $A_6$. On the whole, the present estimations are consistent with those in Ref.~\cite{Zhang:2009kr}. In addition, the authors in Ref.~\cite{Liu:2009dr} also investigate the non-$D\bar{D}$ decays of $\psi(3770)$ by including the meson loop contributions, where only the open channel coupling, i.e., $\psi(3770) D\bar{D}$ has been taken into consideration, and the estimations indicate the meson-loop contributions is important in understanding the non-$D\bar{D}$ branching fraction of $\psi(3770)$.

 \begin{figure*}[t]
	\centering
	\begin{tabular}{ccc}
\includegraphics[width=8.4cm]{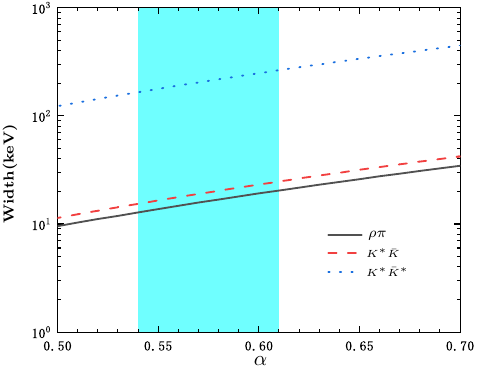} & 
\hspace{0.5cm} &
\includegraphics[width=8.4cm]{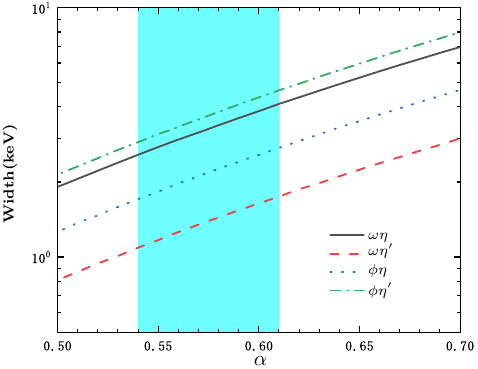}	
	\end{tabular}
	\caption{The partial widths of the light meson decays of  $\psi(3823)$ depending on the model parameter $\alpha$. The vertical band indicates the parameter range determined by $\psi(3770)\to \phi \eta$}\label{Fig:Br2}
\end{figure*}
 
 In Table~\ref{t1}, the measured branching factions of the light meson decays of $\psi(3770)$ are also listed for comparison. From the table one can find that the measured upper limit $\rho^0 \pi^0$, $\omega \eta$, $\phi \eta^\prime$, $K^{\ast+} K^-+c.c.$ are rather small~\cite{CLEO:2005zrs}, which are much smaller than the present estimations and the theoretical investigations in Refs.~\cite{Zhang:2009kr, Liu:2009dr, Li:2013zcr}. It is worth mentioning that most of the branching fractions of the light meson decays of $\psi(3770)$ were reported by CLEO Collaboration in Ref.~\cite{CLEO:2005zrs} with the $e^+e^-$ collision data at $\sqrt{s}=3.773$ GeV ($\mathcal{L}=281~\mathrm{pb}^{-1}$) and $\sqrt{s}=3.671$ GeV ($\mathcal{L}=21~\mathrm{pb}^{-1}$) nearly twenty years ago. Due to the data limitation, the interference effects were neglected in the analysis~\cite{CLEO:2005zrs}. Further experimental analysis with more accurate data may provide more information, which should be accessible by the BESIII Collaboration.

\begin{table}[t]
\centering
\renewcommand\arraystretch{1.5}	
\caption{The partial widths the light meson decays of $\psi(3823)$.\label{t2}}
\begin{tabular}{p{2.5cm}<\centering p{2.5cm}<\centering p{2.5cm}<\centering}
		\toprule[1pt]
		 $\Gamma$(keV)   & Present   & ratio \\
		\midrule[1pt]
		$\phi\eta$               &1.71-2.74               &1.0 \\
        $K^{*}\bar{K^*}$         &166-264                     & 96.3-97.0\\
        $K^{*}\bar{K}$           &15.36-24.71                 &9.0 \\
		$\rho\pi$                &12.81-20.43              &7.5\\
        $\phi\eta^\prime$        &2.89-4.66                 & 1.7\\
		$\omega\eta$             &2.57-4.11               &1.5\\
		$\omega\eta^\prime$      &1.09-1.75             &0.6 \\
		\bottomrule[1pt]
	\end{tabular}
\end{table}

\subsection{light meson decays of $\psi(3823)$}
With the same model parameters, we could estimate the meson loop contributions to the light meson decays of $\psi(3823)$. It should be noted that the width of $\psi(3823)$ has not been well measured, and only the upper limit have been reported by BesIII Collaboration, which is $2.9$ MeV~\cite{BESIII:2022yga}. Thus, in the present estimations, we present the partial widths of light meson decay processes of $\psi(3823)$ depending on the model parameter $\alpha$ in Fig.~\ref{Fig:Br2}. Consider the similarity of the $D$-wave charmonia, we take the same model parameter as the one of the light meson decays of $\psi(3770)$, which is shown as the cyan band in Fig.~\ref{Fig:Br2}. Similar to the case of light meson decay processes $\psi(3770)$, the $\alpha$ dependences of the partial width of the light meson decays of $\psi(3823)$  are almost the same indicating the $\alpha$ independent ratios.

In Table~\ref{t2}, we collect the partial widths of the light meson decays of $\psi(3823)$ estimated with the $\alpha$ range determined by $\psi(3770) \to \phi \eta$. Our estimations indicate that the $K^\ast \bar{K}^\ast$ channel are predominant with the partial width to be about 200 keV. As for  $\rho \pi$ and $K^\ast \bar{K}$ , their partial widths are of order 10 keV, while the partial widths for $\phi \eta$, $\omega \eta$, $\omega \eta^\prime$ are only several keV. Here, we take $\psi(3823) \to \phi \eta $ as scale, the ratios of the partial widths of other processes to that of $\psi(3823)\to \phi \eta$ are also listed in the table.

\begin{figure*}[htb]
	\centering
	\begin{tabular}{ccc}
\includegraphics[width=8.4cm]{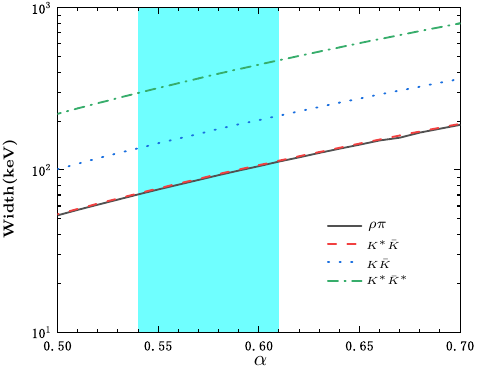} & \hspace{0.5cm} &
\includegraphics[width=8.4cm]{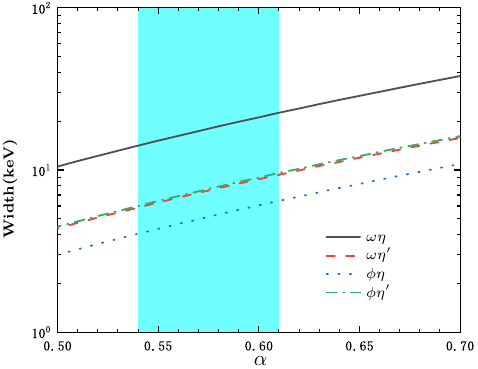}	
	\end{tabular}
	\caption{The same as Fig.~\ref{Fig:Br2} but for $\psi(3842)$.}\label{Fig:Br3}
\end{figure*}

\begin{table}[t]
	\centering
	\renewcommand\arraystretch{1.5}		
	\caption{The partial widths of the light meson decays of $\psi(3842)$.}
	\label{t3}
	\begin{tabular}{p{1.8cm}<\centering p{2cm}<\centering p{2cm}<\centering p{2cm}<\centering}
		\toprule[1pt] 
		$\Gamma$(keV)   & Present    &Ref.~\cite{Bai:2024vul}\footnote{In orde to compare with the present estimations, the partial widths from Ref.~\cite{Bai:2024vul} are evaluated by the branching ratios and the center value of the total width of $\psi(3842)$}            & Ratio \\
		\midrule[1pt]
		$\phi\eta$               &4.04-6.44            &-                & 1.0 \\
        $K^{*}\bar{K^*}$         &298.2-475.0        &0.28-24.36             & 73.8\\
        $K\bar{K}$               &135.9-216.3          &0.56-51.24            & 33.6\\
        $K^{*}\bar{K}$           &71.65-114.2           &0.28-51.24            & 17.7\\
		$\rho \pi$               &70.56-112.46         &1.68-201.88            & 17.5\\
		$\omega\eta$             &14.12-22.49            &0.28-17.92             & 3.5\\
		$\omega\eta^\prime$      &5.87-9.34            &0-1.68              & 1.5\\
		$\phi\eta^\prime$        &6.00-9.57           &-                & 1.5\\
		\bottomrule[1pt]
	\end{tabular}
\end{table}

\subsection{light meson decays of $\psi(3842)$}
In a similar way, we can analysis the light meson decays of the $\psi(3842)$. The partial widths depending on the model parameter $\alpha$ are presented in Fig.~\ref{Fig:Br3}. Although the width of $\psi(3842)$ has been measured to be $(2.8\pm 0.6)$ MeV, we present the estimated partial widths of the light meson decay processes $\psi(3842)$ as that of $\psi(3823)$ and the unobserved $\eta_{c2}$. Similar to the case of $\psi(3770)$ and $\psi(3823)$, the estimated partial widths increase with the $\alpha$ increasing, and the $\alpha$ dependence of these estimated partial widths are similar. 

In Table~\ref{t3}, we collect the partial widths of the considered processes with the $\alpha$ range determined by $\psi(3770) \to \phi \eta$. From the table, one can find that the partial widths of $K^\ast \bar{K}^\ast/K\bar{K}$ and $K^\ast \bar{K}/\rho \pi$ are estimated to be more than and around 100 keV, respectively, the partial width of $\omega \eta$ is about $10\sim 20$ keV, while the partial widths of the rest channel, i.e., $\phi \eta$, $\phi \eta^\prime$ and $\omega \eta^\prime$ are less than 10 keV. Here, we also list the results from Ref.~\cite{Bai:2024vul} for comparison. In Ref.~\cite{Bai:2024vul}, the authors considered the fact that $\psi(3842)$ dominantly decay into $D\bar{D}$, and the $D\bar{D}$ can couples to light meson pairs by exchanging a $D$ or $D^\ast$ meson. In the present estimations, we only considered the coupling $ \psi(3842)\mathcal{D}^\ast \bar{\mathcal{D}}^\ast$ since this coupling is $P$ wave, while $\psi(3842) D\bar{D}$ is $F$ wave, which should be strongly suppressed. Due to coupling $\psi(3842) D_s^{(\ast) +} D_s^{(\ast)-}$, the $\phi \eta$ and $\phi \eta^\prime$ channels are also involved in the present estimations. In addition, with the $\alpha$ range determined by $\psi(3770)\to \phi \eta$, the partial widths are estimated in a narrow range. And in most cases, the partial width estimated in the present work are greater than those in Ref.~\cite{Bai:2024vul}. For example, the partial width of the $K^\ast \bar{K}^\ast$ channel are of at least one order larger than that in Ref.~\cite{Bai:2024vul}. Further experimental measurements for the partial width of $\psi(3842) \to K^\ast \bar{K}^\ast$ could be a good criterion to distinguish the meson loop mechanism involved in the present estimations and in Ref.~\cite{Bai:2024vul}.

\begin{figure}[t]
	\centering
	\includegraphics[width=8.4cm]{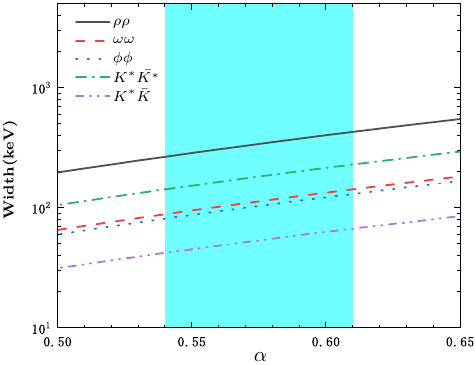}
	\caption{The same as Fig.~\ref{Fig:Br2} but for $\eta_{c2}$.}\label{Fig:Br4}
\end{figure}

\begin{table}[t]
	\centering{}
	\renewcommand\arraystretch{1.5}	
	\caption{The partial widths of the light meson decays of $\eta_{c2}$.}
	\label{t4}
	\begin{tabular}{p{2.6cm}<\centering p{2.6cm}<\centering p{2.6cm}<\centering}
		\toprule[1pt] 
		$\Gamma$(keV)   & Present          &Ratio \\
		\midrule[1pt] 
	   $\phi\phi$                   &81.03-130.40            & 1.0 \\
		$\rho \rho$         &265.64-427.81           & 3.3\\
		$\omega\omega$               &88.34-142.27            & 1.1\\
		$K^{*}\bar{K^*}$             &142.16-228.99         & 1.7\\
		$K^{*}\bar{K}$               &42.07-66.95             & 0.5\\
		\bottomrule[1pt]
	\end{tabular}
\end{table}

\subsection{light meson decays of $\eta_{c2}$}
As for $\eta_{c2}$, it has not been observed experimentally. In the present work, the mass of $\eta_{c2}$ are taken as 3822 MeV, which is determined by the mass gap of $\eta_{c2}$ and $\psi(3823)$ estimated by the quark model~\cite{Barnes:2005pb} and the observed mass of $\psi(3823)$. The partial widths of the light hadron decays of $\eta_{c2}$ are present in Fig.~\ref{Fig:Br4}. Similar to the case of the spin triplet of the $D$ wave charmonia, the partial widths of the involved processes increase with the $\alpha$ increasing. In particular, our estimations indicate that the $\rho \rho$ channel are predominant with the width of $265.64\sim 427.81$ keV in the $\alpha$ range determined by $\psi(3770)\to \phi \eta$. In addition, the widths of $K^\ast \bar{K}^\ast$ is also estimated larger than 100 keV, while the widths for $\omega  \omega$ and $\phi \phi$ are similar, which are of order 100 keV. 

In Table~\ref{t4}, we collect the partial width of the light meson decays of $\eta_{c2}$, where we take $\eta_{c2} \to \phi \phi $ as scale. Our estimations indicate that the width of $\rho \rho$ is about 3 times of $\phi\phi$, and that of $K^\ast \bar{K}^\ast$ is about 1.7 times of the width of $\phi \phi$ channel. From the table, one can find that the partial widths of the light meson decay channels of $\eta_{c2}$ are at least several tens keV, and these light meson decay channels could serve as the important observing channel for $\eta_{c2}$.

\section{Summary}
The light meson decays, as one kind of important non-$D\bar{D}$ decays, are particular interesting since the large branching fraction of the non-$D\bar{D}$ of $\psi(3770)$ reported by the BESIII Collaboration. In the present work, we investigate the light meson decays of $D$-wave charmonia with meson loop mechanism. The model parameter $\alpha$ introduced by the form factor can be determined by the measured branching fraction of $\psi(3770) \to \phi \eta$. With this parameter range, we find that our estimated branching fractions of the light meson decays of $\psi(3770)$ are consistent with the theoretical estimations in the literature, but much greater than the upper limits  reported by the CLEO Collaborations, where interferences effects were neglected in the analysis. More precise measurements with more accurate data may provide further test to the present estimation of the  branching fractions of the light meson decays of $\psi(3770)$, which can be accessed by the BESIII Collaboration.

As for $\psi(3823)$, our estimations indicate that the partial width of $K^\ast \bar{K}^\ast$ are greater than 100 keV, and those for $K^\ast \bar{K}$ and $\rho \pi$ are of order 10 keV, while the partial widths of other channels are only several keV. The present estimation for the partial widths of the light meson decays of $\psi(3842)$ show that the $K^\ast \bar{K}^\ast$ are the predominant decay mode with the partial width of $298.2 \sim 475.0$ keV, and the widths of $K\bar{K}$, $K^\ast \bar{K}+c.c$ and $\rho \pi$ are around 100 keV. Since $\eta_{c2}$ has not been observed experimentally, the mass of $\eta_{c2}$ are taken as $3822$ MeV in the present estimation, which is determined by the mass gap of $\eta_{c2}$ and $\psi(3823)$ estimated by the quark model and the observed mass of $\psi(3823)$. With this mass, we find that $\rho \rho$ channel are the most important light meson decay channel, and the partial width are estimated to be $265.6\sim 472.8$ keV in the considered parameter range. In addition, the partial widths of $K^\ast \bar{K}^\ast$ channel is greater than 100 keV, while those of $\phi \phi$ and $\omega \omega $ are also around 100 keV, which indicates that these light meson decay channels can serve as the observing channels for $\eta_{c2}$.

\section*{ACKNOWLEDGMENTS}
This work is partly supported by the National Natural Science Foundation of China under the Grant Nos. 12175037 and 12335001, as well as supported, in part, by National Key Research and Development Program under the contract No. 2024YFA1610503

\appendix
\section{Amplitudes for $\psi(3770)\to \mathcal{V}\mathcal{V}/\mathcal{P}\mathcal{P}$\label{Sec:Apppsi1}}

\begin{figure}[htb]
	\begin{tabular}{ccc}
		\centering
		\includegraphics[width=2.8cm]{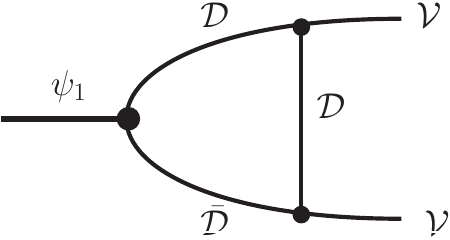}&
		\includegraphics[width=2.8cm]{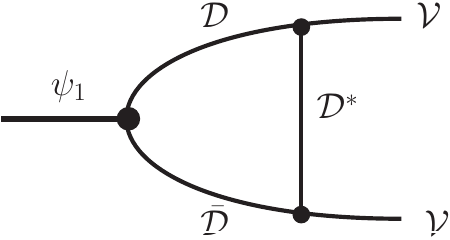}&
		\includegraphics[width=2.8cm]{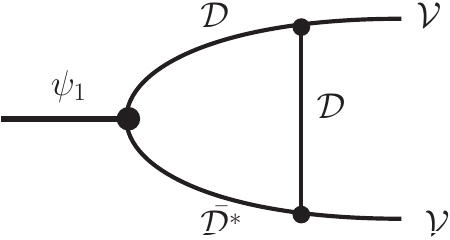}\\
		\\
		$(B_1)$ & $(B_2)$ & $(B_3)$ \\
		\\
		\includegraphics[width=2.8cm]{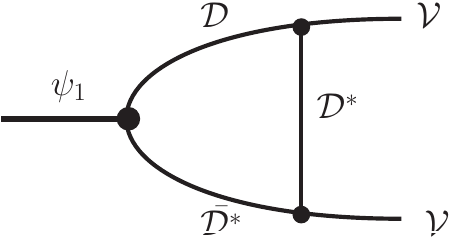}&
		\includegraphics[width=2.8cm]{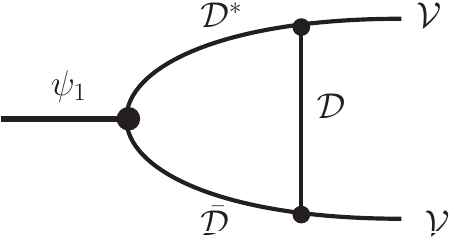}&
		\includegraphics[width=2.8cm]{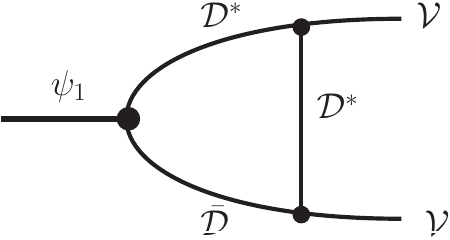}\\
		\\
		$(B_4)$ & $(B_5)$ & $(B_6)$ \\
		\\
		\includegraphics[width=2.8cm]{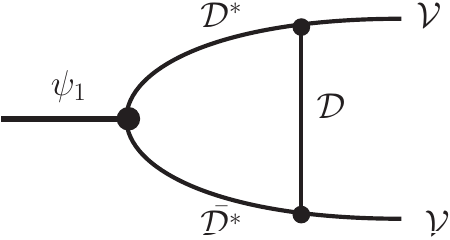}&
		\includegraphics[width=2.8cm]{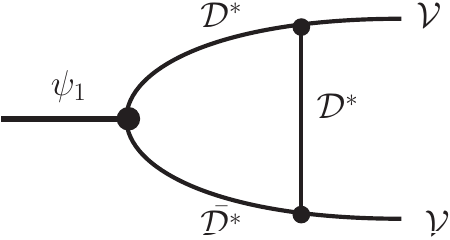}\\
		\\
		$(B_7)$ & $(B_8)$ \\
	\end{tabular}
	\caption{Diagrams contributing to $\psi(3770)\to \mathcal{VV}$ at the hadron level.}\label{Fig:Tri12}
\end{figure}

\begin{figure}[htb]
	\begin{tabular}{ccc}
		\centering
		\includegraphics[width=2.8cm]{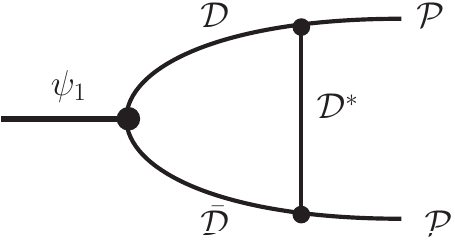}&
		\includegraphics[width=2.8cm]{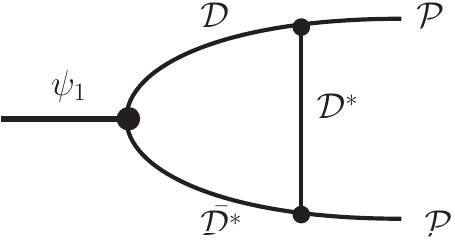}&
		\includegraphics[width=2.8cm]{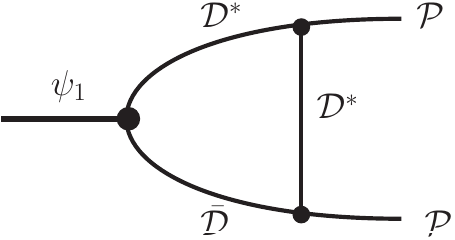}\\
		\\
		$(C_1)$ & $(C_2)$ & $(C_3)$ \\
		\\
		\includegraphics[width=2.8cm]{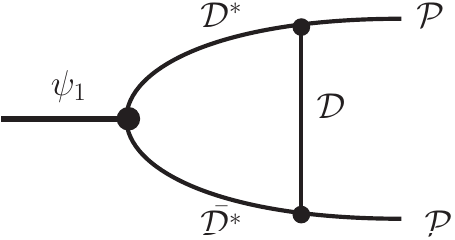}&
		\includegraphics[width=2.8cm]{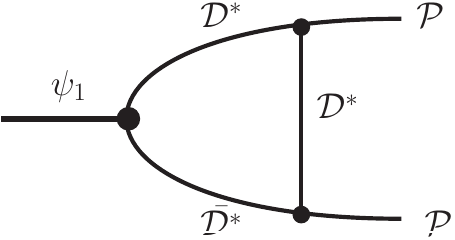}\\
		\\
		$(C_4)$ & $(C_5)$   \\
	\end{tabular}
	\caption{Diagrams contributing to $\psi(3770)\to \mathcal{PP}$ at the hadron level.}\label{Fig:Tri13}
\end{figure}

The amplitudes of $\psi(3770)\to \mathcal{V}\mathcal{V}$ corresponding to Fig.~\ref{Fig:Tri12}-$(B_1)$-$(B_8)$ read,
\begin{eqnarray}	
	\mathcal{M}_{B_1}&=&i^3 \int\frac{d^4 q}{(2\pi)^4}\left[
ig_{\psi_1 \mathcal{D}\mathcal{D}} \epsilon^{\mu}(p)\left(p_{2 \mu}- p_{1 \mu}\right)\right]
\nonumber\\
&&\left[-g_{ \mathcal{D}\mathcal{D}V} \epsilon^{\lambda}\left(p_{3}\right) \left(p_{1\lambda}+q_{\lambda}\right)\right]\left[g_{ \mathcal{D}\mathcal{D}V} \epsilon^{\tau}\left(p_{4}\right) \left(p_{2\tau}-q_{\tau}\right)\right]
\nonumber\\
&&\times\frac{1}{p^2_1-m^2_1}\frac{1}{p^2_2-m^2_2}\frac{1}{q^2-m^2_q}\mathcal{F}^2(m_{q}^2, q^2) ,\nonumber
\end{eqnarray}
\begin{eqnarray}
	\mathcal{M}_{B_2}&=&i^3 \int\frac{d^4 q}{(2\pi)^4}\left[
ig_{\psi_1 \mathcal{D}\mathcal{D}} \epsilon^{\mu}(p)\left(p_{2 \mu}- p_{1 \mu}\right)\right]\nonumber\\
&&\left[2f_{\mathcal{D}^*\mathcal{D}V}\varepsilon_{\chi\lambda\sigma\rho}p_3^\chi\epsilon^\lambda(p_3)\big(q^\sigma+p1^\sigma\big)\right]\nonumber\\
&&\left[2f_{\mathcal{D}^*\mathcal{D}V}\varepsilon_{\kappa\tau\phi\xi}p_4^\kappa\epsilon^\tau(p_4)\big(p_2^\phi-q^\phi\big)\right]\nonumber\\
&&\times\frac{1}{p^2_1-m^2_1}\frac{1}{p^2_2-m^2_2}\frac{-g^{\chi\xi}+q^{\chi} q^{\xi} /m^2_q}{q^2-m^2_q}\mathcal{F}^2(m_{q}^2, q^2), \nonumber\\
\mathcal{M}_{B_3}&=&i^3 \int\frac{d^4 q}{(2\pi)^4}\left[
	g_{\psi_1 \mathcal{D}^{*} \mathcal{D}} \varepsilon_{\mu \nu \alpha \beta}\left(p^{\mu}_{2}-p^{\mu}_{1}\right)  p^{v}\epsilon^{\alpha}\left(p\right)\right]\nonumber\\
	&&\left[-g_{ \mathcal{D}\mathcal{D}V} \epsilon^{\lambda}\left(p_{3}\right) \left(p_{1\lambda}+q_{\lambda}\right)\right]\nonumber\\
	&&\left[2f_{\mathcal{D}^*\mathcal{D}V}\varepsilon_{\kappa\tau\phi\xi}p_4^\kappa\epsilon^\tau(p_4)\big(p_2^\phi-q^\phi\big)\right]\nonumber\\
	&&\times\frac{-g^{\beta\xi}+p_2^{\beta} p_2^{\xi} /m^2_2}{p^2_2-m^2_2}\frac{1}{q^2-m^2_q}\mathcal{F}^2(m_{q}^2 , q^2), \nonumber\\
	\mathcal{M}_{B_4}&=&i^3 \int\frac{d^4 q}{(2\pi)^4}\left[
	g_{\psi_1 \mathcal{D}^{*} \mathcal{D}} \varepsilon_{\mu \nu \alpha \beta}\left(p^{\mu}_{2}-p^{\mu}_{1}\right)  p^{v}\epsilon^{\alpha}\left(p\right)\right]\nonumber\\
	&&\left[-g_{\mathcal{D}^{*} \mathcal{D}P} p_{3\chi }\right]\left[g_{\mathcal{D}^{*} \mathcal{D}^{*}V}g_{\tau \phi} \epsilon^{\tau}\left(p_{4}\right)\left(q^{\xi}-p^{\xi}_{2}\right)\right.\nonumber\\
	&&\left.-4f_{\mathcal{D}^{*} \mathcal{D}^{*}V} \epsilon^{\tau}\left(p_{4}\right)\left(p_4^{\xi}g_{\tau \phi}-p^{\phi}_{4}g_{\tau \xi}\right)\right]\frac{1}{p^2_1-m^2_1}\nonumber\\
	&&\times\frac{-g^{\beta\xi}+p_2^{\beta} p_2^{\xi} /m^2_2}{p^2_2-m^2_2}\frac{-g^{\chi\phi}+q^{\chi} q^{\phi} /m^2_q}{q^2-m^2_q}\mathcal{F}^2(m_{q}^2 , q^2), \nonumber\\
	\mathcal{M}_{B_5}&=&i^3 \int\frac{d^4 q}{(2\pi)^4}\left[
	g_{\psi_{1} \mathcal{D}^{*} \mathcal{D}} \varepsilon_{\mu \nu \alpha \beta} \left(p^{\mu}_{1}-p^{\mu}_{2}\right)  p^{\nu} \epsilon^{\alpha}(p)\right]\nonumber\\
	&&\left[-2f_{\mathcal{D}^*\mathcal{D}V}\varepsilon_{\chi\lambda\sigma\rho}p_3^\chi\epsilon^\lambda(p_3)\big(q^\sigma+p1^\sigma\big)\right]\nonumber\\
	&&\left[g_{ \mathcal{D}\mathcal{D}V} \epsilon^{\tau}\left(p_{4}\right) \left(p_{2\tau}-q_{\tau}\right)\right]\frac{-g^{\beta\phi}+p_1^{\beta} p_1^{\phi} /m^2_1}{p^2_1-m^2_1}
	\nonumber\\
	&&\times\frac{1}{p^2_2-m^2_2}\frac{1}{q^2-m^2_q}\mathcal{F}^2(m_{q}^2, q^2), \nonumber\\
	\mathcal{M}_{B_6}&=&i^3 \int\frac{d^4 q}{(2\pi)^4}\left[
	g_{\psi_{1} \mathcal{D}^{*} \mathcal{D}} \varepsilon_{\mu \nu \alpha \beta} \left(p^{\mu}_{1}-p^{\mu}_{2}\right)  p^{\nu} \epsilon^{\alpha}(p)\right]\nonumber\\
	&&\left[g_{\mathcal{D}^{*} \mathcal{D}^{*}V}g_{\lambda \sigma} \epsilon^{\lambda}\left(p_{3}\right)\left(q^{\rho}+p^{\rho}_{1}\right)\right.\nonumber\\
	&&\left.+4f_{\mathcal{D}^{*} \mathcal{D}^{*}V} \epsilon^{\lambda}\left(p_{4}\right)\left(p_3^{\rho}g_{\lambda \sigma}-p^{\sigma}_{3}g_{\lambda \rho}\right)\right]\nonumber\\&&
\left[2f_{\mathcal{D}^*\mathcal{D}V}\varepsilon_{\kappa\tau\phi\xi}p_4^\kappa\epsilon^\tau(p_4)\big(p_2^\phi-q^\phi\big)\right]\frac{1}{p^2_2-m^2_2}
	\nonumber\\
	&&\times\frac{-g^{\beta\rho}+p_1^{\beta} p_1^{\rho} /m^2_1}{p^2_1-m^2_1}\frac{-g^{\sigma\xi}+q^{\sigma} q^{\xi} /m^2_q}{q^2-m^2_q}\mathcal{F}^2(m_{q}^2, q^2), \nonumber\\
	\mathcal{M}_{B_7}&=&i^3 \int\frac{d^4 q}{(2\pi)^4}\left[ig_{\psi_{1} \mathcal{D}^{*} \mathcal{D}^{*}}\epsilon^{\mu}(p)\left(-4 ( p_{1\mu}-p_{2\mu})  g^{\alpha \beta}\right.\right.\nonumber\\
	&&\left.\left.+p^{\alpha}_{1}g_{\mu}^{\beta}-p^{\beta}_{2}g^{\alpha}_{\mu}\right) \right] \left[-2f_{\mathcal{D}^*\mathcal{D}V}\varepsilon_{\chi\lambda\sigma\rho}p_3^\chi\epsilon^\lambda(p_3)\big(q^\sigma+p1^\sigma\big)\right]\nonumber\\
	&&\left[2f_{\mathcal{D}^*\mathcal{D}V}\varepsilon_{\kappa\tau\phi\xi}p_4^\kappa\epsilon^\tau(p_4)\big(q^\phi-p2^\phi\big)\right]
	\nonumber\\
	&&\times\frac{-g^{\beta\rho}+p_1^{\beta} p_1^{\rho} /m^2_1}{p^2_1-m^2_1}\frac{-g^{\alpha\xi}+p_2^{\alpha} p_2^{\xi} /m^2_2}{p^2_2-m^2_2}\nonumber\\
	&&\times\frac{1}{q^2-m^2_q}\mathcal{F}^2(m_{q}^2, q^2), \nonumber
\end{eqnarray}
\begin{eqnarray}
	\mathcal{M}_{B_8}&=&i^3 \int\frac{d^4 q}{(2\pi)^4}\left[ig_{\psi_{1} \mathcal{D}^{*} \mathcal{D}^{*}}\epsilon^{\mu}(p)\left(-4 ( p_{1\mu}-p_{2\mu})  g^{\alpha \beta}\right.\right.\nonumber\\
	&&\left.\left.+p^{\alpha}_{1}g_{\mu}^{\beta}-p^{\beta}_{2}g^{\alpha}_{\mu}\right) \right] \left[g_{\mathcal{D}^{*} \mathcal{D}^{*}V}g_{\lambda \sigma} 
	\epsilon^{\lambda}\left(p_{3}\right)\left(q^{\rho}+p^{\rho}_{1}\right)\right.\nonumber\\
	&&\left.+4f_{\mathcal{D}^{*} \mathcal{D}^{*}V} \epsilon^{\lambda}\left(p_{4}\right)\left(p_3^{\rho}g_{\lambda \sigma}-p^{\sigma}_{3}g_{\lambda \rho}\right)\right]\nonumber\\
	&&\left[g_{\mathcal{D}^{*} \mathcal{D}^{*}V}g_{\tau \phi} \epsilon^{\tau}\left(p_{4}\right)\left(q^{\xi}-p^{\xi}_{2}\right)\right.\nonumber\\
	&&\left.-4f_{\mathcal{D}^{*} \mathcal{D}^{*}V} \epsilon^{\tau}\left(p_{4}\right)\left(p_4^{\xi}g_{\tau \phi}-p^{\phi}_{4}g_{\tau \xi}\right)\right]\nonumber\\
	&&\times\frac{-g^{\beta\rho}+p_1^{\beta} p_1^{\rho} /m^2_1}{p^2_1-m^2_1}\frac{-g^{\alpha\xi}+p_2^{\alpha} p_2^{\xi} /m^2_2}{p^2_2-m^2_2}\nonumber\\
	&&\times\frac{-g^{\sigma\phi}+q^{\sigma} q^{\phi} /m^2_q}{q^2-m^2_q}\mathcal{F}^2(m_{q}^2, q^2).
\end{eqnarray}
Then the total amplitude for $\psi(3770) \to \mathcal{V}\mathcal{V}$ read,
\begin{eqnarray}
	\mathcal{M}_{\psi_1\to \mathcal{V}\mathcal{V}} =\sum_{j=1}^{8}\mathcal{M}_{B_j}.
\end{eqnarray}	

The amplitudes of $\psi(3770)\to \mathcal{P}\mathcal{P}$ corresponding to Fig.~\ref{Fig:Tri13}-$(C_{1})$-$(C_{5})$ read,
\begin{eqnarray}
	\mathcal{M}_{C_{1}}&=&i^3 \int\frac{d^4 q}{(2\pi)^4}\left[
	ig_{\psi_1 \mathcal{D}\mathcal{D}} \epsilon^{\mu}(p)\left(p_{2 \mu}- p_{1 \mu}\right)\right]
	\nonumber\\
	&&\left[g_{ \mathcal{D}^{*} \mathcal{D}P} p_{3\chi }\right]\left[g_{ \mathcal{D}^{*} \mathcal{D}P} p_{4\kappa}\right]
	\nonumber\\
	&&\times\frac{1}{p^2_1-m^2_1}\frac{1}{p^2_2-m^2_2}\frac{-g^{\chi\kappa}+q^{\chi} q^{\kappa} /m^2_q}{q^2-m^2_q}\mathcal{F}^2(m_{q}^2, q^2), \nonumber\\
	\mathcal{M}_{C_{2}}&=&i^3 \int\frac{d^4 q}{(2\pi)^4}\left[
	g_{\psi_1 \mathcal{D}^{*} \mathcal{D}} \varepsilon_{\mu \nu \alpha \beta}\left(p^{\mu}_{2}-p^{\mu}_{1}\right)  p^{v}\epsilon^{\alpha}\left(p\right)\right]\nonumber\\
	&&\left[g_{ \mathcal{D}^{*} \mathcal{D}P} p_{3\chi }\right]\left[\frac{1}{2} g_{ \mathcal{D}^{*} \mathcal{D}^{*}P } \varepsilon_{\kappa\tau\phi\xi}  p_{4}^{\tau} \left(q^{\phi}-q^{\phi}_{2}\right) \right]\nonumber\\
	&&\left.-4f_{\mathcal{D}^{*} \mathcal{D}^{*}V} \epsilon^{\tau}\left(p_{4}\right)\left(p_4^{\xi}g_{\tau \phi}-p^{\phi}_{4}g_{\tau \xi}\right)\right]\frac{1}{p^2_1-m^2_1}\nonumber\\
	&&\times\frac{-g^{\beta\xi}+p_2^{\beta} p_2^{\xi} /m^2_2}{p^2_2-m^2_2}\frac{-g^{\chi\kappa}+q^{\chi} q^{\kappa} /m^2_q}{q^2-m^2_q}\mathcal{F}^2(m_{q}^2 , q^2), \nonumber\\
	\mathcal{M}_{C_{3}}&=&i^3 \int\frac{d^4 q}{(2\pi)^4}\left[
	g_{\psi_{1} \mathcal{D}^{*} \mathcal{D}} \varepsilon_{\mu \nu \alpha \beta} \left(p^{\mu}_{1}-p^{\mu}_{2}\right)  p^{\nu} \epsilon^{\alpha}(p)\right]\nonumber\\
	&&\left[\frac{1}{2} g_{ \mathcal{D}^{*} \mathcal{D}^{*}P } \varepsilon_{\chi\lambda\sigma\rho}  p_{3}^{\lambda} \left(p^{\sigma}_{1}+q^{\sigma}\right) \right]\left[g_{ \mathcal{D}^{*} \mathcal{D}P} p_{4\kappa}\right]
	\nonumber\\
	&&\times\frac{-g^{\beta\chi}+p_1^{\beta} p_1^{\chi} /m^2_1}{p^2_1-m^2_1}\nonumber\\
	&&\times\frac{1}{p^2_2-m^2_2}\frac{-g^{\rho\kappa}+q^{\rho} q^{\kappa} /m^2_q}{q^2-m^2_q}\mathcal{F}^2(m_{q}^2, q^2), \nonumber
\end{eqnarray}
\begin{eqnarray}	
	\mathcal{M}_{C_{4}}&=&i^3 \int\frac{d^4 q}{(2\pi)^4}\left[ig_{\psi_{1} \mathcal{D}^{*} \mathcal{D}^{*}}\epsilon^{\mu}(p)\left(-4 ( p_{1\mu}-p_{2\mu})  g^{\alpha \beta}\right.\right.\nonumber\\
	&&\left.\left.+p^{\alpha}_{1}g_{\mu}^{\beta}-p^{\beta}_{2}g^{\alpha}_{\mu}\right) \right]\left[-g_{ \mathcal{D}^{*} \mathcal{D}P} p_{3\chi }\right] \nonumber\\
	&&\left[-g_{ \mathcal{D}^{*} \mathcal{D}P} p_{4\kappa }\right]\frac{1}{p^2_2-m^2_2}\frac{-g^{\beta\chi}+p_1^{\beta} p_1^{\chi} /m^2_1}{p^2_1-m^2_1}
	\nonumber\\
	&&\times\frac{-g^{\alpha\xi}+q^{\alpha} q^{\xi} /m^2_q}{q^2-m^2_q}\mathcal{F}^2(m_{q}^2, q^2), \nonumber\\
	\mathcal{M}_{C_{5}}&=&i^3 \int\frac{d^4 q}{(2\pi)^4}\left[ig_{\psi_{1} \mathcal{D}^{*} \mathcal{D}^{*}}\epsilon^{\mu}(p)\left(-4 ( p_{1\mu}-p_{2\mu})  g^{\alpha \beta}\right.\right.\nonumber\\
	&&\left.\left.+p^{\alpha}_{1}g_{\mu}^{\beta}-p^{\beta}_{2}g^{\alpha}_{\mu}\right) \right]\left[\frac{1}{2} g_{ \mathcal{D}^{*} \mathcal{D}^{*}P } \varepsilon_{\chi\lambda\sigma\rho}  p_{3}^{\lambda} \left(p^{\sigma}_{1}+q^{\sigma}\right) \right]\nonumber\\
	&&\left[\frac{1}{2} g_{ \mathcal{D}^{*} \mathcal{D}^{*}P } \varepsilon_{\kappa\tau\phi\xi}  p_{4}^{\tau} \left(q^{\phi}-q^{\phi}_{2}\right) \right]
	\nonumber\\
	&&\times\frac{-g^{\beta\chi}+p_1^{\beta} p_1^{\chi} /m^2_1}{p^2_1-m^2_1}\frac{-g^{\alpha\xi}+p_2^{\alpha} p_2^{\xi} /m^2_2}{p^2_2-m^2_2}\nonumber\\
	&&\times\frac{-g^{\rho\kappa}+q^{\rho} q^{\kappa} /m^2_q}{q^2-m^2_q}\mathcal{F}^2(m_{q}^2, q^2).
\end{eqnarray}

Then the total amplitude for $\psi(3770) \to \mathcal{P}\mathcal{P}$ read,
\begin{eqnarray}
	\mathcal{M}_{\psi_1\to \mathcal{V}\mathcal{V}} =\sum_{j=1}^{5}\mathcal{M}_{C_j}.
\end{eqnarray}

\section{Amplitudes for $\psi(3823)\to \mathcal{P}\mathcal{V}/\mathcal{V}\mathcal{V}/\mathcal{P}\mathcal{P}$\label{Sec:Apppsi2}}

\begin{figure}[t]
	\begin{tabular}{ccc}
		\centering
		\includegraphics[width=2.8cm]{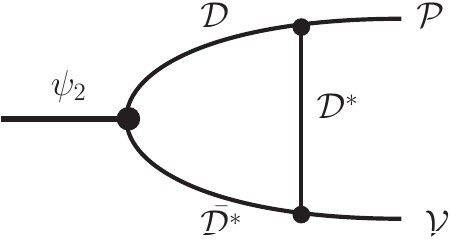}&
		\includegraphics[width=2.8cm]{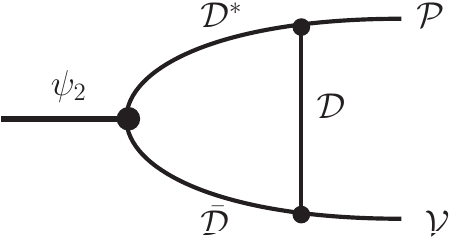}&
		\includegraphics[width=2.8cm]{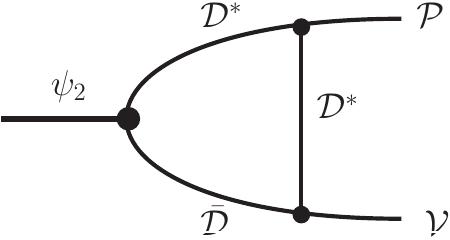}\\
		\\
		$(D_1)$ & $(D_2)$ & $(D_3)$ \\
		\\
		\includegraphics[width=2.8cm]{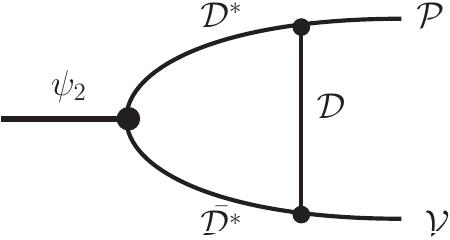}&
		\includegraphics[width=2.8cm]{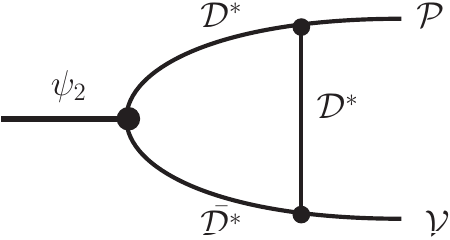}\\
		\\
		$(D_4)$ & $(D_5)$  \\
	\end{tabular}
	\caption{Diagrams contributing to $\psi(3823)\to \mathcal{PV}$ at the hadron level .}\label{Fig:Tri21}
\end{figure}

The amplitudes of $\psi_2(3823)\to \mathcal{P}\mathcal{V}$ corresponding to Fig.~\ref{Fig:Tri21}-$(D_1)$-$(D_5)$ read,	

\begin{eqnarray}
	\mathcal{M}_{D_1}&=&i^3 \int\frac{d^4 q}{(2\pi)^4}
	\left[g_{\psi_2 \mathcal{D}^{*} \mathcal{D}} \epsilon^{\mu\nu}(p)(p_{1 \nu}- p_{2 \nu})\right]\nonumber\\&&
	\left[g_{ \mathcal{D}^{*} \mathcal{D}P} p_{3\chi }\right]\left[g_{\mathcal{D}^{*} \mathcal{D}^{*}V}g_{\tau \phi} \epsilon^{\tau}\left(p_{4}\right)\left(q^{\xi}-p^{\xi}_{2}\right)\right.\nonumber\\
	&&\left.-4f_{\mathcal{D}^{*} \mathcal{D}^{*}V} \epsilon^{\tau}\left(p_{4}\right)\left(p_4^{\xi}g_{\tau \phi}-p^{\phi}_{4}g_{\tau \xi}\right)\right]\nonumber\\
	&&\times\frac{1}{p^2_1-m^2_1}\frac{-g^{\mu\xi}+p_2^{\mu} p_2^{\xi} /m^2_2}{p^2_2-m^2_2}	\nonumber\\
	&& \times\frac{-g^{\chi\phi}+q^{\chi} q^{\phi} /m^2_q}{q^2-m^2_q}\mathcal{F}^2(m_q^2, q^2),\nonumber\\
	\mathcal{M}_{D_2}&=&i^3 \int\frac{d^4 q}{(2\pi)^4}\left[
	g_{\psi_2 \mathcal{D}^{*} \mathcal{D}} \epsilon^{\mu\nu}(p)(p_{2 \nu}- p_{1 \nu})\right]
\left[-g_{ \mathcal{D}^{*} \mathcal{D}P} p_{3\chi }\right]\nonumber\\
	&&\left[g_{ \mathcal{D}\mathcal{D}V} \epsilon^{\tau}\left(p_{4}\right) \left(p_{2\tau}-q_{\tau}\right)\right]\frac{-g^{\mu\chi}+p_1^{\mu} p_1^{\chi} /m^2_1}{p^2_1-m^2_1}
	\nonumber\\
	&&\times\frac{1}{p^2_2-m^2_2}\frac{1}{q^2-m^2_q}\mathcal{F}^2(m_q^2,q^2), \nonumber\\
	\mathcal{M}_{D_3}&=&i^3 \int\frac{d^4 q}{(2\pi)^4}\left[
	g_{\psi_2 \mathcal{D}^{*} \mathcal{D}} \epsilon^{\mu\nu}(p)\left(p_{2 \nu}- p_{1 \nu}\right)\right]\nonumber\\
	&&\left[\frac{1}{2} g_{ \mathcal{D}^{*} \mathcal{D}^{*}P } \varepsilon_{\chi\lambda\sigma\rho}  p_{3}^{\lambda} \left(p^{\sigma}_{1}+q^{\sigma}\right) \right]\nonumber\\
	&& \left[2f_{\mathcal{D}^*\mathcal{D}V}\varepsilon_{\kappa\tau\phi\xi}p_4^\kappa\epsilon^\tau(p_4)\big(p_2^\phi-q^\phi\big)\right]\nonumber\\
	&&\times\frac{-g^{\mu\rho}+p_1^{\mu} p_1^{\rho} /m^2_1}{p^2_1-m^2_1}\frac{1}{p^2_2-m^2_2}\nonumber\\
	&&\times\frac{-g^{\chi\xi}+q^{\chi} q^{\xi} /m^2_q}{q^2-m^2_q}\mathcal{F}^2(m_q^2,q^2),\nonumber\\
	\mathcal{M}_{D_4}&=&i^3 \int\frac{d^4 q}{(2\pi)^4}
	\Big[ig_{\psi_2 \mathcal{D}^{*} \mathcal{D}^{*}} \varepsilon_{\alpha\beta\mu\nu}p^{\mu} \epsilon^{\alpha\gamma}(p) \Big((p_{2} ^{\beta}- p_{1}^{\beta})\nonumber\\
	&&g_{\delta}^{\nu}g_{\theta}^{\gamma}+(p_{2}^{\beta}- p_{1}^{\beta})g_{\gamma\delta}g_{\nu}^{\theta}\Big)\Big] \left[-g_{ \mathcal{D}^{*} \mathcal{D}P} p_{3\chi }\right]\nonumber\\
	&&\Big[2f_{\mathcal{D}^*\mathcal{D}V}\varepsilon_{\kappa\tau\phi\xi}p_4^\kappa\epsilon^\tau(p_4)\big(q^\phi-p_2^\phi\big)\Big]\frac{1}{q^2-m^2_q}\nonumber\\ 
	&&\times \frac{-g^{\theta\chi}+p_1^{\theta} p_1^{\chi} /m^2_1}{p^2_1-m^2_1} \frac{-g^{\delta\xi}+p_2^{\delta} p_2^{\xi} /m^2_2}{p^2_2-m^2_2}\mathcal{F}^2(m_q^2, q^2),\nonumber\\
	\mathcal{M}_{D_5}&=&i^3 \int\frac{d^4 q}{(2\pi)^4}
	\left[ig_{\psi_2 \mathcal{D}^{*} \mathcal{D}^{*}} \varepsilon_{\alpha\beta\mu\nu}p^{\mu} \epsilon^{\alpha\gamma}(p) \Big((p_{2} ^{\beta}- p_{1}^{\beta})g_{\delta}^{\nu}g_{\theta}^{\gamma}\right.\nonumber\\
	&&\left.+(p_{2}^{\beta}- p_{1}^{\beta})g_{\gamma\delta}g_{\nu}^{\theta}\Big)\right]\left[\frac{1}{2} g_{ \mathcal{D}^{*} \mathcal{D}^{*}P } \varepsilon_{\chi\lambda\sigma\rho}  p_{3}^{\lambda} \left(p^{\sigma}_{1}+q^{\sigma}\right) \right]\nonumber\\
	&& \left[g_{\mathcal{D}^{*} \mathcal{D}^{*}V}g_{\tau \phi} \epsilon^{\tau}\left(p_{4}\right)\left(q^{\xi}-p^{\xi}_{2}\right)\right.\nonumber\\
	&&\left.-4f_{\mathcal{D}^{*} \mathcal{D}^{*}V} \epsilon^{\tau}\left(p_{4}\right)\left(p_4^{\xi}g_{\tau \phi}-p^{\phi}_{4}g_{\tau \xi}\right)\right]\nonumber\\
	&&\times\frac{-g^{\theta\rho}+p_1^{\theta} p_1^{\rho} /m^2_1}{p^2_1-m^2_1}\frac{-g^{\delta\xi}+p_2^{\delta} p_2^{\xi} /m^2_2}{p^2_2-m^2_2} \nonumber\\
	&&\times\frac{-g^{\chi\phi}+q^{\chi} q^{\phi} /m^2_q}{q^2-m^2_q}\mathcal{F}^2(m_q^2,q^2).
\end{eqnarray}
Then the total amplitude for $\psi(3823) \to \mathcal{P}\mathcal{V}$ read,
\begin{eqnarray}
	\mathcal{M}_{\psi_2\to \mathcal{P}\mathcal{V}} =\sum_{j=1}^{5}\mathcal{M}_{D_j}.
\end{eqnarray}	

\begin{figure}[t]
	\begin{tabular}{ccc}
		\centering
		\includegraphics[width=2.8cm]{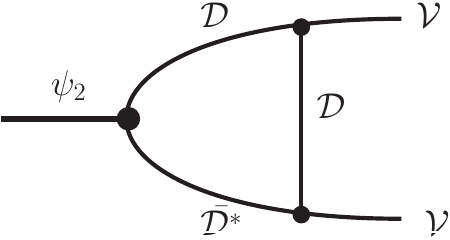}&
		\includegraphics[width=2.8cm]{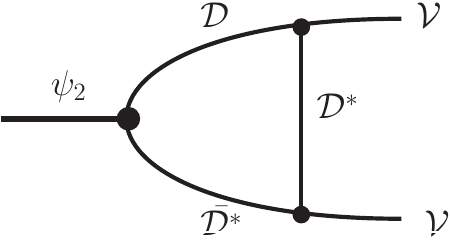}&
		\includegraphics[width=2.8cm]{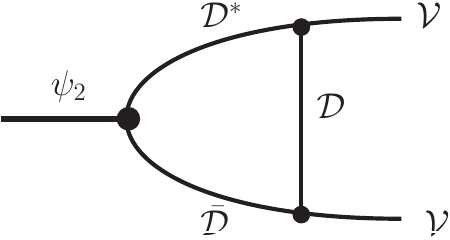}\\
		\\
		$(E_1)$ & $(E_2)$ & $(E_3)$ \\
		\\
		\includegraphics[width=2.8cm]{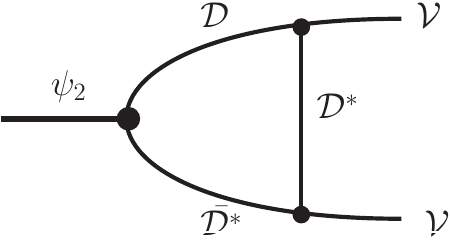}&
		\includegraphics[width=2.8cm]{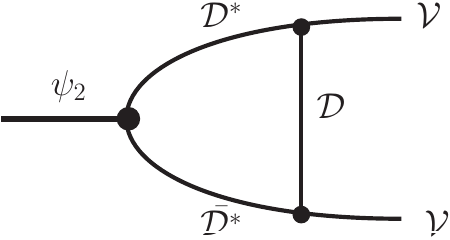}&
		\includegraphics[width=2.8cm]{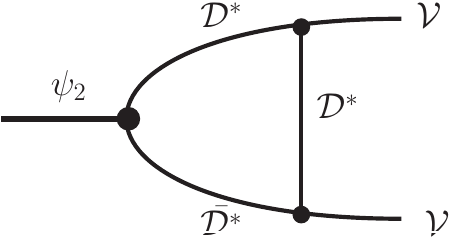}\\
		\\
		$(E_4)$ & $(E_5)$ & $(E_6)$ \\
	\end{tabular}
	\caption{Diagrams contributing to $\psi(3823)\to \mathcal{VV}$ at the hadron level.}\label{Fig:Tri22}
\end{figure}

The amplitudes of $\psi_2(3823)\to \mathcal{V}\mathcal{V}$  corresponding to Fig.~\ref{Fig:Tri22}-$(E_1)$-$(E_6)$ read,	

\begin{eqnarray}
	\mathcal{M}_{E_1}&=&i^3 \int\frac{d^4 q}{(2\pi)^4}
	\left[g_{\psi_2 \mathcal{D}^{*} \mathcal{D}} \epsilon^{\mu\nu}(p)(p_{1 \nu}- p_{2 \nu})\right]\nonumber\\&&
	\left[-g_{ \mathcal{D}\mathcal{D}V} \epsilon^{\lambda}\left(p_{3}\right) \left(p_{1\lambda}+q_{\lambda}\right)\right]\nonumber\\
	&&\left[2f_{\mathcal{D}^*\mathcal{D}V}\varepsilon_{\kappa\tau\phi\xi}p_4^\kappa\epsilon^\tau(p_4)\big(q^\phi-p2^\phi\big)\right]\nonumber\\
	&&\times\frac{1}{p^2_1-m^2_1}\frac{-g^{\mu\xi}+p_2^{\mu} p_2^{\xi} /m^2_2}{p^2_2-m^2_2}	\nonumber\\
	&& \times\frac{1}{q^2-m^2_q}\mathcal{F}^2(m_q^2, q^2),\nonumber\\
	\mathcal{M}_{E_2}&=&i^3 \int\frac{d^4 q}{(2\pi)^4}
	\left[g_{\psi_2 \mathcal{D}^{*} \mathcal{D}} \epsilon^{\mu\nu}(p)(p_{1 \nu}- p_{2 \nu})\right]\nonumber\\&&
\left[2f_{D^*DV}\varepsilon_{\chi\lambda\sigma\rho}p_3^\chi\epsilon^\lambda(p_3)\big(q^\sigma+p1^\sigma\big)\right]\nonumber\\
	&&\left[g_{\mathcal{D}^{*} \mathcal{D}^{*}V}g_{\tau \phi} \epsilon^{\tau}\left(p_{4}\right)\left(q^{\xi}-p^{\xi}_{2}\right)\right.\nonumber\\
	&&\left.-4f_{\mathcal{D}^{*} \mathcal{D}^{*}V} \epsilon^{\tau}\left(p_{4}\right)\left(p_4^{\xi}g_{\tau \phi}-p^{\phi}_{4}g_{\tau \xi}\right)\right]
	\nonumber\\
	&&\frac{1}{p^2_1-m^2_1}\frac{-g^{\mu\xi}+p_2^{\mu} p_2^{\xi} /m^2_2}{p^2_2-m^2_2}	\nonumber\\
	&& \times\frac{-g^{\rho\phi}+q^{\rho} q^{\phi} /m^2_q}{q^2-m^2_q}\mathcal{F}^2(m_q^2, q^2),\nonumber\\
	\mathcal{M}_{E_3}&=&i^3 \int\frac{d^4 q}{(2\pi)^4}\left[
	g_{\psi_2 \mathcal{D}^{*} \mathcal{D}} \epsilon^{\mu\nu}(p)(p_{2 \nu}- p_{1 \nu})\right]\nonumber\\&&
	\left[-2f_{\mathcal{D}^*\mathcal{D}V}\varepsilon_{\chi\lambda\sigma\rho}p_3^\chi\epsilon^\lambda(p_3)\big(q^\sigma+p1^\sigma\big)\right]\nonumber\\
	&&\left[g_{ \mathcal{D}\mathcal{D}V} \epsilon^{\tau}\left(p_{4}\right) \left(p_{2\tau}-q_{\tau}\right)\right]\frac{-g^{\mu\rho}+p_1^{\mu} p_1^{\rho} /m^2_1}{p^2_1-m^2_1}
	\nonumber\\
	&&\times\frac{1}{p^2_2-m^2_2}\frac{1}{q^2-m^2_q}\mathcal{F}^2(m_q^2,q^2), \nonumber
\end{eqnarray}
\begin{eqnarray}
	\mathcal{M}_{E_4}&=&i^3 \int\frac{d^4 q}{(2\pi)^4}\left[
	g_{\psi_2 \mathcal{D}^{*} \mathcal{D}} \epsilon^{\mu\nu}(p)\left(p_{2 \nu}- p_{1 \nu}\right)\right]
	\nonumber\\	&&\left[2f_{\mathcal{D}^*\mathcal{D}V}\varepsilon_{\chi\lambda\sigma\rho}p_3^\chi\epsilon^\lambda(p_3)\big(q^\sigma+p1^\sigma\big)\right]\nonumber\\&&\Big[g_{\mathcal{D}^{*} \mathcal{D}^{*}V}g_{\tau \phi} \epsilon^{\tau}\left(p_{4}\right)\left(q^{\xi}-p^{\xi}_{2}\right)\nonumber\\
	&&-4f_{\mathcal{D}^{*} \mathcal{D}^{*}V} \epsilon^{\tau}\left(p_{4}\right)\left(p_4^{\xi}g_{\tau \phi}-p^{\phi}_{4}g_{\tau \xi}\right)\Big]\nonumber\\
	&&\times\frac{-g^{\mu\rho}+p_1^{\mu} p_1^{\rho} /m^2_1}{p^2_1-m^2_1}\frac{1}{p^2_2-m^2_2}\frac{-g^{\phi\xi}+q^{\phi} q^{\xi} /m^2_q}{q^2-m^2_q}\mathcal{F}^2(m_q^2,q^2), \nonumber\\
	\mathcal{M}_{E_5}&=&i^3 \int\frac{d^4 q}{(2\pi)^4}
	\Big[ig_{\psi_2 \mathcal{D}^{*} \mathcal{D}^{*}} \varepsilon_{\alpha\beta\mu\nu}p^{\mu} \epsilon^{\alpha\gamma}(p) \Big((p_{2} ^{\beta}- p_{1}^{\beta})g_{\delta}^{\nu}g_{\theta}^{\gamma}\nonumber\\
	&&+(p_{2}^{\beta}- p_{1}^{\beta})g_{\gamma\delta}g_{\nu}^{\theta}\Big)\Big] \left[-2f_{\mathcal{D}^*\mathcal{D}V}\varepsilon_{\chi\lambda\sigma\rho}p_3^\chi\epsilon^\lambda(p_3)\big(q^\sigma+p1^\sigma\big)\right]\nonumber\\
	&&\Big[2f_{\mathcal{D}^*\mathcal{D}V}\varepsilon_{\kappa\tau\phi\xi}p_4^\kappa\epsilon^\tau(p_4)\big(q^\phi-p_2^\phi\big)\Big]\frac{1}{q^2-m^2_q}\nonumber\\ 
	&&\times \frac{-g^{\theta\rho}+p_1^{\theta} p_1^{\rho} /m^2_1}{p^2_1-m^2_1} \frac{-g^{\delta\xi}+p_2^{\delta} p_2^{\xi} /m^2_2}{p^2_2-m^2_2}\mathcal{F}^2(m_q^2, q^2),\nonumber\\
	\mathcal{M}_{E_6}&=&i^3 \int\frac{d^4 q}{(2\pi)^4}
	\left[ig_{\psi_2 \mathcal{D}^{*} \mathcal{D}^{*}} \varepsilon_{\alpha\beta\mu\nu}p^{\mu} \epsilon^{\alpha\gamma}(p) \Big((p_{2} ^{\beta}- p_{1}^{\beta})g_{\delta}^{\nu}g_{\theta}^{\gamma}\right.\nonumber\\
	&&\left.+(p_{2}^{\beta}- p_{1}^{\beta})g_{\gamma\delta}g_{\nu}^{\theta}\Big)\right]\left[g_{\mathcal{D}^{*} \mathcal{D}^{*}V}g_{\lambda \sigma} \epsilon^{\lambda}\left(p_{3}\right)\left(q^{\rho}+p^{\rho}_{1}\right)\right.\nonumber\\
	&&\left.+4f_{\mathcal{D}^{*} \mathcal{D}^{*}V} \epsilon^{\lambda}\left(p_{4}\right)\left(p_3^{\rho}g_{\lambda \sigma}-p^{\sigma}_{3}g_{\lambda \rho}\right)\right]\Big[g_{\mathcal{D}^{*} \mathcal{D}^{*}V}g_{\tau \phi}\nonumber\\
	&& \epsilon^{\tau}\left(p_{4}\right)\left(q^{\xi}-p^{\xi}_{2}\right)-4f_{\mathcal{D}^{*} \mathcal{D}^{*}V} \epsilon^{\tau}\left(p_{4}\right)\left(p_4^{\xi}g_{\tau \phi}-p^{\phi}_{4}g_{\tau \xi}\right)\Big]\nonumber\\
	&&\times\frac{-g^{\theta\rho}+p_1^{\theta} p_1^{\rho} /m^2_1}{p^2_1-m^2_1}\frac{-g^{\sigma\xi}+p_2^{\sigma} p_2^{\xi} /m^2_2}{p^2_2-m^2_2}\nonumber\\
	&&\times\frac{-g^{\delta\phi}+q^{\delta} q^{\phi} /m^2_q}{q^2-m^2_q}\mathcal{F}^2(m_{q}^2, q^2).
\end{eqnarray}
Then the total amplitude for $\psi(3823) \to \mathcal{V}\mathcal{V}$ read,
\begin{eqnarray}
	\mathcal{M}_{\psi_2\to \mathcal{V}\mathcal{V}} =\sum_{j=1}^{6}\mathcal{M}_{E_j}.
\end{eqnarray}

\begin{figure}[t]
	\begin{tabular}{cc}
		\centering
		\includegraphics[width=2.8cm]{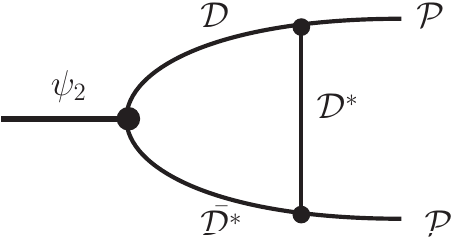}&
		\includegraphics[width=2.8cm]{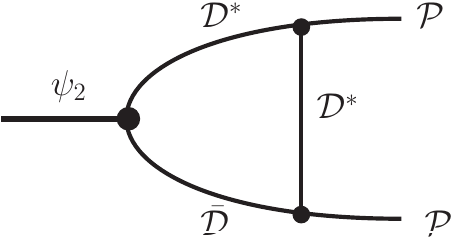}\\
		\\
		$(F_1)$ & $(F_2)$\\
		\\
		\includegraphics[width=2.8cm]{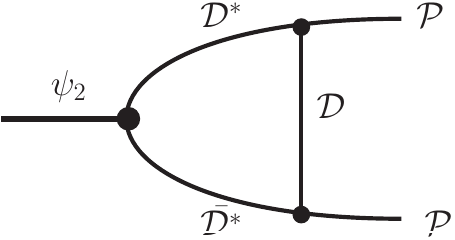}&
		\includegraphics[width=2.8cm]{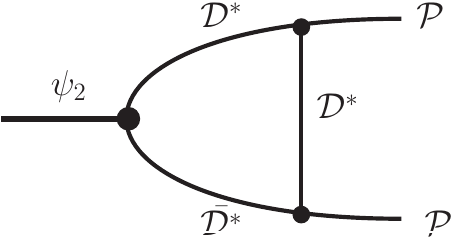}\\
		\\
		$(F_3)$ & $(F_4)$  \\
	\end{tabular}
	\caption{Diagrams contributing to $\psi(3823)\to \mathcal{PP}$ at the hadron level.}\label{Fig:Tri23}
\end{figure}

The amplitudes of $\psi_2(3823)\to \mathcal{P}\mathcal{P}$  corresponding to Fig.~\ref{Fig:Tri23}-$(F_1)$-$(F_4)$ read,	
\begin{eqnarray}
	\mathcal{M}_{F_1}&=&i^3 \int\frac{d^4 q}{(2\pi)^4}
	\left[g_{\psi_2 \mathcal{D}^{*} \mathcal{D}} \epsilon^{\mu\nu}(p)(p_{1 \nu}- p_{2 \nu})\right]\nonumber\\
	&&\left[g_{ \mathcal{D}^{*} \mathcal{D}P} p_{3\chi }\right]\left[\frac{1}{2} g_{ \mathcal{D}^{*} \mathcal{D}^{*}P } \varepsilon_{\kappa\tau\phi\xi}  p_{4}^{\tau} \left(q^{\phi}-q^{\phi}_{2}\right) \right]	\nonumber\\
	&&\times\frac{1}{p^2_1-m^2_1}\frac{-g^{\mu\xi}+p_2^{\mu} p_2^{\xi} /m^2_2}{p^2_2-m^2_2}	\nonumber\\
	&& \times\frac{-g^{\chi\kappa}+q^{\chi} q^{\kappa} /m^2_q}{q^2-m^2_q}\mathcal{F}^2(m_q^2, q^2),\nonumber\\
	\mathcal{M}_{F_2}&=&i^3 \int\frac{d^4 q}{(2\pi)^4}\left[
	g_{\psi_2 \mathcal{D}^{*} \mathcal{D}} \epsilon^{\mu\nu}(p)(p_{2 \nu}- p_{1 \nu})\right]\nonumber\\
    &&\left[\frac{1}{2} g_{ \mathcal{D}^{*} \mathcal{D}^{*}P } \varepsilon_{\chi\lambda\sigma\rho}  p_{3}^{\lambda} \left(p^{\sigma}_{1}+q^{\sigma}\right) \right]\nonumber\\
	&&\left[g_{ \mathcal{D}^{*} \mathcal{D}P} p_{4\kappa}\right]\frac{-g^{\mu\chi}+p_1^{\mu} p_1^{\chi} /m^2_1}{p^2_1-m^2_1}
	\nonumber\\
	&&\times\frac{1}{p^2_2-m^2_2}\frac{-g^{\rho\kappa}+q^{\rho} q^{\kappa} /m^2_q}{q^2-m^2_q}\mathcal{F}^2(m_q^2,q^2),\nonumber\\
	\mathcal{M}_{F_3}&=&i^3 \int\frac{d^4 q}{(2\pi)^4}
	\Big[ig_{\psi_2 \mathcal{D}^{*} \mathcal{D}^{*}} \varepsilon_{\alpha\beta\mu\nu}p^{\mu} \epsilon^{\alpha\gamma}(p) \nonumber\\
	&&\Big((p_{2} ^{\beta}- p_{1}^{\beta})g_{\delta}^{\nu}g_{\theta}^{\gamma}+(p_{2}^{\beta}- p_{1}^{\beta})g_{\gamma\delta}g_{\nu}^{\theta}\Big)\Big]\nonumber\\
	&& \left[-g_{ \mathcal{D}^{*} \mathcal{D}P} p_{3\chi }\right]\left[-g_{ \mathcal{D}^{*} \mathcal{D}P} p_{4\kappa }\right]\frac{1}{q^2-m^2_q}\nonumber\\ 
	&&\times \frac{-g^{\theta\chi}+p_1^{\theta} p_1^{\chi} /m^2_1}{p^2_1-m^2_1} \frac{-g^{\delta\kappa}+p_2^{\delta} p_2^{\kappa} /m^2_2}{p^2_2-m^2_2}\mathcal{F}^2(m_q^2, q^2),\nonumber\\
	\mathcal{M}_{F_4}&=&i^3 \int\frac{d^4 q}{(2\pi)^4}
	\left[ig_{\psi_2 \mathcal{D}^{*} \mathcal{D}^{*}} \varepsilon_{\alpha\beta\mu\nu}p^{\mu} \epsilon^{\alpha\gamma}(p) \right.\nonumber\\
	&&\left.\Big((p_{2} ^{\beta}- p_{1}^{\beta})g_{\delta}^{\nu}g_{\theta}^{\gamma}+(p_{2}^{\beta}- p_{1}^{\beta})g_{\gamma\delta}g_{\nu}^{\theta}\Big)\right]\nonumber\\
	&&\left[\frac{1}{2} g_{ \mathcal{D}^{*} \mathcal{D}^{*}P } \varepsilon_{\chi\lambda\sigma\rho}  p_{3}^{\lambda} \left(p^{\sigma}_{1}+q^{\sigma}\right) \right]\nonumber\\
	&&\left[\frac{1}{2} g_{ \mathcal{D}^{*} \mathcal{D}^{*}P } \varepsilon_{\kappa\tau\phi\xi}  p_{4}^{\tau} \left(q^{\phi}-q^{\phi}_{2}\right) \right]\frac{-g^{\theta\rho}+p_1^{\theta} p_1^{\rho} /m^2_1}{p^2_1-m^2_1}\nonumber\\
	&&\times\frac{-g^{\delta\xi}+p_2^{\delta} p_2^{\xi} /m^2_2}{p^2_2-m^2_2}\frac{-g^{\chi\kappa}+q^{\chi} q^{\kappa} /m^2_q}{q^2-m^2_q}\mathcal{F}^2(m_q^2,q^2).\nonumber\\
\end{eqnarray}

Then the total amplitude for $\psi(3823) \to \mathcal{P}\mathcal{P}$ read,
\begin{eqnarray}
	\mathcal{M}_{\psi_2\to \mathcal{P}\mathcal{P}} =\sum_{j=1}^{4}\mathcal{M}_{F_j}.
\end{eqnarray}

\section{Amplitudes for $\psi(3842)\to \mathcal{V}\mathcal{P}/\mathcal{V}\mathcal{V}/\mathcal{P}\mathcal{P}$\label{Sec:Apppsi3}}

\begin{figure}[t]
	\begin{tabular}{ccc}
		\centering
		\includegraphics[width=2.8cm]{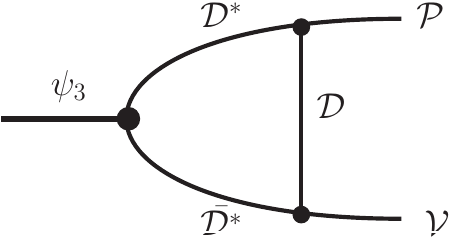}&
		\includegraphics[width=2.8cm]{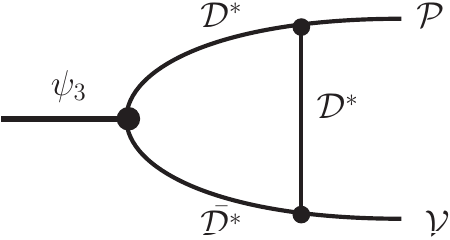}&
		\\
		$(G_1)$ & $(G_2)$  \\
		\\
		
	\end{tabular}
	\caption{Diagrams contributing to $\psi(3842)\to \mathcal{PV}$ at the hadron level .}\label{Fig:Tri31}
\end{figure}

The amplitudes of $\psi(3842)\to \mathcal{V}\mathcal{P}$ corresponding to Fig.~\ref{Fig:Tri31}-$(G_1)$-$(G_2)$ read,
\begin{eqnarray}
	\mathcal{M}_{G_1}&=&i^3 \int\frac{d^4 q}{(2\pi)^4}
	\Big[ig_{\psi_3 \mathcal{D}^{*} \mathcal{D}^{*}} \epsilon^{\mu\nu\alpha}(p)\left((p_{2 \mu}- p_{1 \mu})g_{\alpha}^{\theta}g_{\nu}^{\delta}\right.\nonumber\\
	&&\left.+(p_{2 \mu}- p_{1 \mu})g_{\nu}^{\theta}g_{\alpha}^{\delta}\right)\Big]\left[-g_{ \mathcal{D}^{*} \mathcal{D}P} p_{3\chi }\right]\nonumber\\
	&&\Big[2f_{\mathcal{D}^*\mathcal{D}V}\varepsilon_{\kappa\tau\phi\xi}p_4^\kappa\epsilon^\tau(p_4)\big(q^\phi-p_2^\phi\big)\Big]\nonumber\\
	&&\times\frac{1}{q^2-m^2_q} \frac{-g^{\theta\chi}+p_1^{\theta} p_1^{\chi} /m^2_1}{p^2_1-m^2_1}\nonumber\\
	&&\times\frac{-g^{\delta\xi}+p_2^{\delta} p_2^{\xi} /m^2_2}{p^2_2-m^2_2}\mathcal{F}^2(m_{q}, q^2), \nonumber\\
	\mathcal{M}_{G_2}&=&i^3 \int\frac{d^4 q}{(2\pi)^4}
	\Big[ig_{\psi_3 \mathcal{D}^{*} \mathcal{D}^{*}} \epsilon^{\mu\nu\alpha}(p)\left((p_{2 \mu}- p_{1 \mu})g_{\alpha}^{\theta}g_{\nu}^{\delta}\right.\nonumber\\
	&&\left.+(p_{2 \mu}- p_{1 \mu})g_{\nu}^{\theta}g_{\alpha}^{\delta}\right)\Big]\left[\frac{1}{2} g_{ \mathcal{D}^{*} \mathcal{D}^{*}P } \varepsilon_{\chi\lambda\sigma\rho}  p_{3}^{\lambda} \left(p^{\sigma}_{1}+q^{\sigma}\right) \right]\nonumber\\
	&&\left[g_{\mathcal{D}^{*} \mathcal{D}^{*}V}g_{\tau \phi} \epsilon^{\tau}\left(p_{4}\right)\left(q^{\xi}-p^{\xi}_{2}\right)-4f_{\mathcal{D}^{*} \mathcal{D}^{*}V} \epsilon^{\tau}\left(p_{4}\right)\left(p_4^{\xi}g_{\tau \phi}\right.\right.\nonumber\\
	&&\left.\left.-p^{\phi}_{4}g_{\tau \xi}\right)\right]\frac{-g^{\theta\rho}+p_1^{\theta} p_1^{\rho} /m^2_1}{p^2_1-m^2_1}\frac{-g^{\delta\xi}+p_2^{\delta} p_2^{\xi} /m^2_2}{p^2_2-m^2_2}\nonumber\\
	&&\times\frac{-g^{\chi\phi}+q^{\chi} q^{\phi} /m^2_q}{q^2-m^2_q}\mathcal{F}^2(m_{q}, q^2).
\end{eqnarray}
Then the total amplitude for $\psi(3842) \to \mathcal{P}\mathcal{V} $ reads
\begin{eqnarray}
	\mathcal{M}_{\psi_3 \to\mathcal{P}\mathcal{V}} =\mathcal{M}_{G_1} +\mathcal{M}_{G_2}.
\end{eqnarray}

\begin{figure}[t]
	\begin{tabular}{cc}
		\centering
		\includegraphics[width=2.8cm]{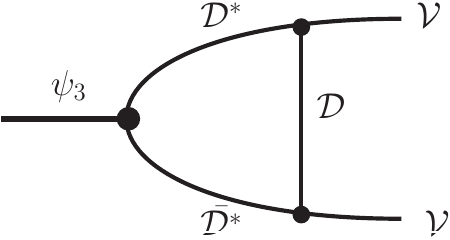}&
		\includegraphics[width=2.8cm]{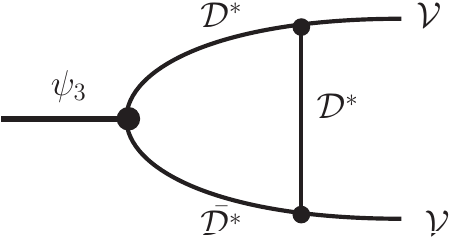}\\
		\\
		$(H_1)$ & $(H_2)$ \\
	\end{tabular}
	\caption{Diagrams contributing to $\psi(3842)\to \mathcal{VV}$ at the hadron level.}\label{Fig:Tri32}
\end{figure}

The amplitudes corresponding to Fig.~\ref{Fig:Tri32}-$(H_1)$-$(H_2)$ about $\psi(3842)\to \mathcal{V}\mathcal{V}$ read,

\begin{eqnarray}
	\mathcal{M}_{H_1}&=&i^3 \int\frac{d^4 q}{(2\pi)^4}
	\Big[ig_{\psi_3 \mathcal{D}^{*} \mathcal{D}^{*}} \epsilon^{\mu\nu\alpha}(p)\left((p_{2 \mu}- p_{1 \mu})g_{\alpha}^{\theta}g_{\nu}^{\delta}\right.\nonumber\\
	&&\left.+(p_{2 \mu}- p_{1 \mu})g_{\nu}^{\theta}g_{\alpha}^{\delta}\right)\Big]\left[-2f_{\mathcal{D}^*\mathcal{D}V}\varepsilon_{\chi\lambda\sigma\rho}p_3^\chi\epsilon^\lambda(p_3)\right.\nonumber\\	&&\left.\big(q^\sigma+p1^\sigma\big)\right]\Big[2f_{\mathcal{D}^*\mathcal{D}V}\varepsilon_{\kappa\tau\phi\xi}p_4^\kappa\epsilon^\tau(p_4)\big(q^\phi-p_2^\phi\big)\Big]\nonumber\\
	&&\times\frac{1}{q^2-m^2_q} \frac{-g^{\theta\chi}+p_1^{\theta} p_1^{\chi} /m^2_1}{p^2_1-m^2_1}\nonumber\\
	&&\times\frac{-g^{\delta\phi}+p_2^{\delta} p_2^{\phi} /m^2_2}{p^2_2-m^2_2}\mathcal{F}^2(m_{q}, q^2)\nonumber\\
	\mathcal{M}_{H_2}&=&i^3 \int\frac{d^4 q}{(2\pi)^4}
	\Big[ig_{\psi_3 \mathcal{D}^{*} \mathcal{D}^{*}} \epsilon^{\mu\nu\alpha}(p)\left((p_{2 \mu}- p_{1 \mu})g_{\alpha}^{\theta}g_{\nu}^{\delta}\right.\nonumber\\
	&&\left.+(p_{2 \mu}- p_{1 \mu})g_{\nu}^{\theta}g_{\alpha}^{\delta}\right)\Big]\left[g_{\mathcal{D}^{*} \mathcal{D}^{*}V}g_{\lambda \sigma} 
	\epsilon^{\lambda}\left(p_{3}\right)\right.\nonumber\\
	&&\left.\left(q^{\rho}+p^{\rho}_{1}\right)+4f_{\mathcal{D}^{*} \mathcal{D}^{*}V} \epsilon^{\lambda}\left(p_{4}\right)\left(p_3^{\rho}g_{\lambda \sigma}-p^{\sigma}_{3}g_{\lambda \rho}\right)\right]\nonumber\\
	&&\left[g_{\mathcal{D}^{*} \mathcal{D}^{*}V}g_{\tau \phi} \epsilon^{\tau}\left(p_{4}\right)\left(q^{\xi}-p^{\xi}_{2}\right)\right.\nonumber\\
	&&\left.-4f_{\mathcal{D}^{*} \mathcal{D}^{*}V} \epsilon^{\tau}\left(p_{4}\right)\left(p_4^{\xi}g_{\tau \phi}-p^{\phi}_{4}g_{\tau \xi}\right)\right]\nonumber\\
	&&\frac{-g^{\theta\rho}+p_1^{\theta} p_1^{\rho} /m^2_1}{p^2_1-m^2_1}\frac{-g^{\delta\xi}+p_2^{\delta} p_2^{\xi} /m^2_2}{p^2_2-m^2_2}\nonumber\\
	&&\times\frac{-g^{\sigma\phi}+q^{\sigma} q^{\phi} /m^2_q}{q^2-m^2_q}\mathcal{F}^2(m_{q}, q^2).
\end{eqnarray}
Then the total amplitude for $\psi(3842) \to \mathcal{V}\mathcal{V} $ reads
\begin{eqnarray}
	\mathcal{M}_{\psi_3 \to\mathcal{V}\mathcal{V}} =\mathcal{M}_{H_1} +\mathcal{M}_{H_2}.
\end{eqnarray}

\begin{figure}[t]
	\begin{tabular}{cc}
		\centering
		\includegraphics[width=2.8cm]{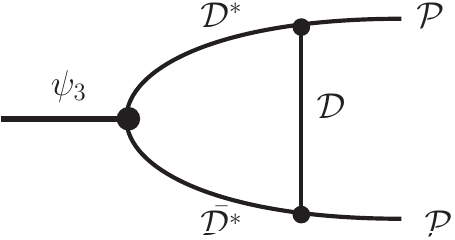}&
		\includegraphics[width=2.8cm]{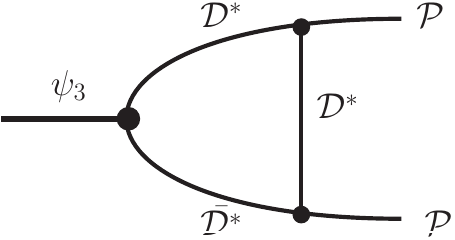}\\
		\\
		$(I_1)$ & $(I_2)$ \\
	\end{tabular}
	\caption{Diagrams contributing to $\psi(3842)\to \mathcal{PP}$ at the hadron level.}\label{Fig:Tri33}
\end{figure}
The amplitudes corresponding to Fig.~\ref{Fig:Tri33}-$(I_1)$-$(I_2)$ about $\psi(3842)\to \mathcal{V}\mathcal{V}$ read,
\begin{eqnarray}
	\mathcal{M}_{I_1}&=&i^3 \int\frac{d^4 q}{(2\pi)^4}
	\Big[ig_{\psi_3 \mathcal{D}^{*} \mathcal{D}^{*}} \epsilon^{\mu\nu\alpha}(p)\left((p_{2 \mu}- p_{1 \mu})g_{\alpha}^{\theta}g_{\nu}^{\delta}\right.\nonumber\\
	&&\left.+(p_{2 \mu}- p_{1 \mu})g_{\nu}^{\theta}g_{\alpha}^{\delta}\right)\Big]\left[-g_{ \mathcal{D}^{*} \mathcal{D}P} p_{3\chi }\right]\left[-g_{ \mathcal{D}^{*} \mathcal{D}P} p_{4\kappa }\right]\nonumber\\
	&&\times\frac{1}{q^2-m^2_q} \frac{-g^{\theta\chi}+p_1^{\theta} p_1^{\chi} /m^2_1}{p^2_1-m^2_1}\nonumber\\
	&&\times\frac{-g^{\delta\kappa}+p_2^{\delta} p_2^{\kappa} /m^2_2}{p^2_2-m^2_2}\mathcal{F}^2(m_{q}, q^2),\nonumber
\end{eqnarray}
\begin{eqnarray}
	\mathcal{M}_{I_2}&=&i^3 \int\frac{d^4 q}{(2\pi)^4}
	\Big[ig_{\psi_3 \mathcal{D}^{*} \mathcal{D}^{*}} \epsilon^{\mu\nu\alpha}(p)\left((p_{2 \mu}- p_{1 \mu})g_{\alpha}^{\theta}g_{\nu}^{\delta}\right.\nonumber\\
	&&\left.+(p_{2 \mu}- p_{1 \mu})g_{\nu}^{\theta}g_{\alpha}^{\delta}\right)\Big]\left[\frac{1}{2} g_{ \mathcal{D}^{*} \mathcal{D}^{*}P } \varepsilon_{\chi\lambda\sigma\rho}  p_{3}^{\lambda} \left(p^{\sigma}_{1}+q^{\sigma}\right) \right]\nonumber\\
	&&\left[\frac{1}{2} g_{ \mathcal{D}^{*} \mathcal{D}^{*}P } \varepsilon_{\kappa\tau\phi\xi}  p_{4}^{\tau} \left(q^{\phi}-q^{\phi}_{2}\right) \right]\nonumber\\
	&&\times\frac{-g^{\theta\rho}+p_1^{\theta} p_1^{\rho} /m^2_1}{p^2_1-m^2_1}\frac{-g^{\delta\xi}+p_2^{\delta} p_2^{\xi} /m^2_2}{p^2_2-m^2_2}\nonumber\\
	&&\times\frac{-g^{\chi\kappa}+q^{\chi} q^{\kappa} /m^2_q}{q^2-m^2_q}\mathcal{F}^2(m_{q}, q^2).
\end{eqnarray}
Then the total amplitude for $\psi(3842) \to \mathcal{P}\mathcal{P} $ reads
\begin{eqnarray}
	\mathcal{M}_{\psi_3 \to\mathcal{P}\mathcal{P}} =\mathcal{M}_{I_1} +\mathcal{M}_{I_2}.
\end{eqnarray}

\section{Amplitudes for $\eta_{c2}\to \mathcal{V}\mathcal{V}/\mathcal{P}\mathcal{V}$\label{Sec:Appetac2}}
\begin{figure}[t]
	\begin{tabular}{ccc}
		\centering
		\includegraphics[width=2.8cm]{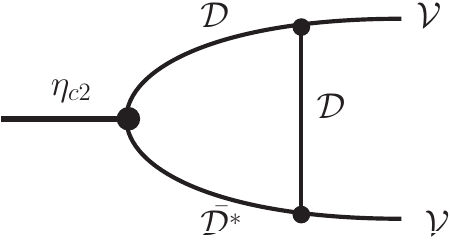}&
		\includegraphics[width=2.8cm]{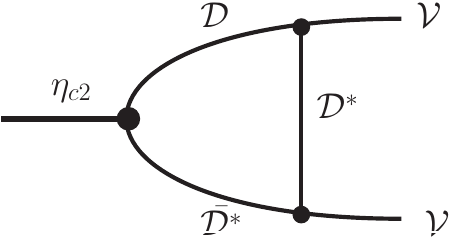}&
		\includegraphics[width=2.8cm]{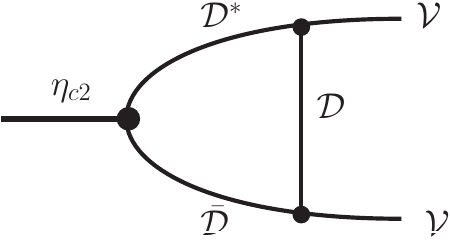}\\
		\\
		$(J_1)$ & $(J_2)$ & $(J_3)$ \\
		\\
		\includegraphics[width=2.8cm]{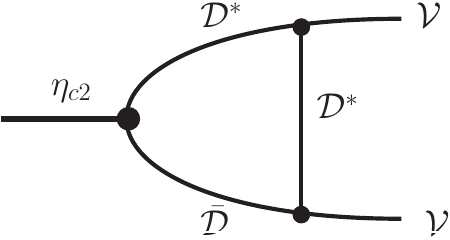}&
		\includegraphics[width=2.8cm]{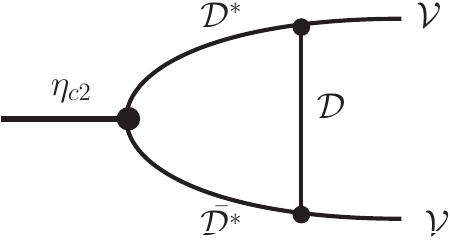}&
		\includegraphics[width=2.8cm]{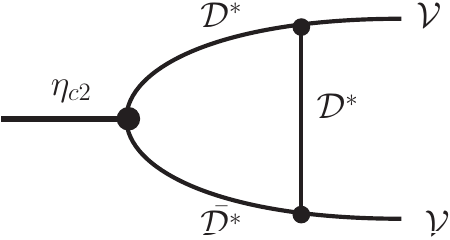}\\
		\\
		$(J_4)$ & $(J_5)$ & $(J_6)$ \\
	\end{tabular}
	\caption{Diagrams contributing to $\eta_{c2} \to \mathcal{VV}$ at the hadron level .}\label{Fig:Tri41}
\end{figure}

The amplitudes of the process $\eta_{c2}\to \mathcal{V}\mathcal{V}$, corresponding to Fig.~\ref{Fig:Tri41}-$(J_1)$-$(J_6)$, reads, 

\begin{eqnarray}
	\mathcal{M}_{J_1}&=&i^3 \int\frac{d^4 q}{(2\pi)^4}
	\Big[ig_{\eta_{c2} \mathcal{D}^{*} \mathcal{D}} \varepsilon^{\mu\alpha}(p)\left(p_{1\mu}- p_{2\mu}\right)\Big]\nonumber\\
	&&\left[-g_{ \mathcal{D}\mathcal{D}V} \epsilon^{\lambda}\left(p_{3}\right) \left(p_{1\lambda}+q_{\lambda}\right)\right]\nonumber\\
	&&\Big[2f_{\mathcal{D}^*\mathcal{D}V}\varepsilon_{\kappa\tau\phi\xi}p_4^\kappa\epsilon^\tau(p_4)\big(q^\phi-p_2^\phi\big)\Big]\frac{1}{p^2_1-m^2_1} \nonumber\\
	&& \times\frac{-g^{\alpha\xi}+p_2^{\alpha} p_2^{\xi} /m^2_2}{p^2_2-m^2_2}\frac{1}{q^2-m^2_q}\mathcal{F}^2(m_q^2,q^2),\nonumber
\end{eqnarray}
\begin{eqnarray}	
	\mathcal{M}_{J_2}&=&i^3 \int\frac{d^4 q}{(2\pi)^4}
	\Big[ig_{\eta_{c2} \mathcal{D}^{*} \mathcal{D}} \varepsilon^{\mu\alpha}(p)\left(p_{1\mu}- p_{2\mu}\right)\Big]\nonumber\\
	&&\left[2f_{\mathcal{D}^*\mathcal{D}V}\varepsilon_{\chi\lambda\sigma\rho}p_3^\chi\epsilon^\lambda(p_3)\big(q^\sigma+p1^\sigma\big)\right]\left[g_{\mathcal{D}^{*} \mathcal{D}^{*}V}g_{\tau \phi} \epsilon^{\tau}\right.\nonumber\\
	&&\left.\left(p_{4}\right)\left(q^{\xi}-p^{\xi}_{2}\right)-4f_{\mathcal{D}^{*} \mathcal{D}^{*}V} \epsilon^{\tau}\left(p_{4}\right)\left(p_4^{\xi}g_{\tau \phi}-p^{\phi}_{4}g_{\tau \xi}\right)\right] \nonumber\\
	&& \times\frac{1}{p^2_1-m^2_1}\frac{-g^{\alpha\xi}+p_2^{\alpha} p_2^{\xi} /m^2_2}{p^2_2-m^2_2}\nonumber\\
	&& \times\frac{-g^{\rho\phi}+q^{\rho} q^{\phi} /m^2_q}{q^2-m^2_q}\mathcal{F}^2(m_q^2,q^2),\nonumber \\
	\mathcal{M}_{J_3}&=&i^3 \int\frac{d^4 q}{(2\pi)^4}
	\Big[ig_{\eta_{c2} \mathcal{D}^{*} \mathcal{D}} \varepsilon^{\mu\alpha}(p)\left(p_{1\mu}- p_{2\mu}\right)\Big] \nonumber\\
	&&\left[-2f_{\mathcal{D}^*\mathcal{D}V}\varepsilon_{\chi\lambda\sigma\rho}p_3^\chi\epsilon^\lambda(p_3)\big(q^\sigma+p1^\sigma\big)\right] \nonumber\\
	&& \Big[g_{ \mathcal{D}\mathcal{D}V} \epsilon^{\tau}\left(p_{4}\right) \left(p_{2\tau}-q_{\tau}\right) \Big] \nonumber\\
	&&\times\frac{-g^{\alpha\rho}+p_1^{\alpha} p_1^{\rho} /m^1_2}{p^2_1-m^2_1}\frac{1}{p^2_2-m^2_2} \nonumber\\
	&&\times\frac{1}{q^2-m^2_q}\mathcal{F}^2(m_q^2,q^2),\nonumber\\
	\mathcal{M}_{J_4}&=&i^3 \int\frac{d^4 q}{(2\pi)^4}
	\Big[ig_{\eta_{c2} \mathcal{D}^{*} \mathcal{D}} \varepsilon^{\mu\alpha}(p)\left(p_{1\mu}- p_{2\mu}\right)\Big] \nonumber\\
	&&\left[g_{\mathcal{D}^{*} \mathcal{D}^{*}V}g_{\lambda \sigma} 
	\epsilon^{\lambda}\left(p_{3}\right)\left(q^{\rho}+p^{\rho}_{1}\right)\right.\nonumber\\
	&&\left.+4f_{\mathcal{D}^{*} \mathcal{D}^{*}V} \epsilon^{\lambda}\left(p_{4}\right)\left(p_3^{\rho}g_{\lambda \sigma}-p^{\sigma}_{3}g_{\lambda \rho}\right)\right]\nonumber\\
	&& \Big[2f_{\mathcal{D}^*\mathcal{D}V}\varepsilon_{\kappa\tau\phi\xi}p_4^\kappa\epsilon^\tau(p_4)\big(p_2^\phi-q^\phi\big)\Big] \nonumber\\
	&&\times\frac{-g^{\alpha\sigma}+p_1^{\alpha} p_1^{\sigma} /m^2_1}{p^2_1-m^2_1} \frac{1}{p^2_2-m^2_2}\nonumber\\
	&&\times\frac{-g^{\rho\xi}+q^{\rho} q^{\xi} /m^2_q}{q^2-m^2_q}\mathcal{F}^2(m_q^2,q^2),\nonumber\\ 
	\mathcal{M}_{J_5}&=&i^3 \int\frac{d^4 q}{(2\pi)^4}
	\Big[ig_{\eta_{c2} \mathcal{D}^{*} \mathcal{D}^{*}} \varepsilon_{\alpha\beta\nu\gamma}p^{\gamma} \epsilon^{\mu\nu}(p)(p_{2\mu}- p_{1\mu})\Big] \nonumber\\
	&&\left[-2f_{\mathcal{D}^*\mathcal{D}V}\varepsilon_{\chi\lambda\sigma\rho}p_3^\chi\epsilon^\lambda(p_3)\big(q^\sigma+p1^\sigma\big)\right]\nonumber\\
	&&\Big[2f_{\mathcal{D}^*\mathcal{D}V}\varepsilon_{\kappa\tau\phi\xi}p_4^\kappa\epsilon^\tau(p_4)\big(q^\phi-p_2^\phi\big)]\nonumber\\
	&&\times\frac{-g^{\alpha\rho}+p_1^{\alpha} p_1^{\rho} /m^2_1}{p^2_1-m^2_1} \frac{-g^{\beta\xi}+p_2^{\beta} p_2^{\xi} /m^2_2}{p^2_2-m^2_2}\nonumber\\
	&&\times\frac{1}{q^2-m^2_q}\mathcal{F}^2(m_q^2,q^2),\nonumber\\
	%
	\mathcal{M}_{J_6}&=&i^3 \int\frac{d^4 q}{(2\pi)^4}
	\Big[ig_{\eta_{c2} \mathcal{D}^{*} \mathcal{D}^{*}} \varepsilon_{\alpha\beta\nu\gamma}p^{\gamma} \epsilon^{\mu\nu}(p)(p_{2\mu}- p_{1\mu})\Big] \nonumber\\
	&&
	\left[g_{\mathcal{D}^{*} \mathcal{D}^{*}V}g_{\lambda \sigma} 
	\epsilon^{\lambda}\left(p_{3}\right)\left(q^{\rho}+p^{\rho}_{1}\right)\right.\nonumber\\
	&&\left.+4f_{\mathcal{D}^{*} \mathcal{D}^{*}V} \epsilon^{\lambda}\left(p_{4}\right)\left(p_3^{\rho}g_{\lambda \sigma}-p^{\sigma}_{3}g_{\lambda \rho}\right)\right]\nonumber\\
	&&\Big[g_{\mathcal{D}^{*} \mathcal{D}^{*}V}g_{\tau \phi} \epsilon^{\tau}\left(p_{4}\right)\left(q^{\xi}-p^{\xi}_{2}\right)\nonumber\\
	&&-4f_{\mathcal{D}^{*} \mathcal{D}^{*}V} \epsilon^{\tau}\left(p_{4}\right)\left(p_4^{\xi}g_{\tau \phi}-p^{\phi}_{4}g_{\tau \xi}\right)\Big] \nonumber\\
	&&\times\frac{-g^{\alpha\sigma}+p_1^{\alpha} p_1^{\sigma} /m^2_1}{p^2_1-m^2_1}\frac{-g^{\beta\xi}+p_2^{\beta} p_2^{\xi} /m^2_2}{p^2_2-m^2_2}\nonumber\\
	&&\times\frac{-g^{\rho\phi}+q^{\rho} q^{\phi} /m^2_q}{q^2-m^2_q}\mathcal{F}^2(m_q^2,q^2).\nonumber\\
\end{eqnarray}

Finally, the total amplitude for $\eta_{c2} \to  \mathcal{V}\mathcal{V}$  reads,
\begin{eqnarray}
	\mathcal{M}_{\eta_{c2}\to \mathcal{V}\mathcal{V}} =\sum_{j=1}^{6}\mathcal{M}_{J_j}.
\end{eqnarray}

\begin{figure}[t]
	\begin{tabular}{ccc}
		\centering
		\includegraphics[width=2.8cm]{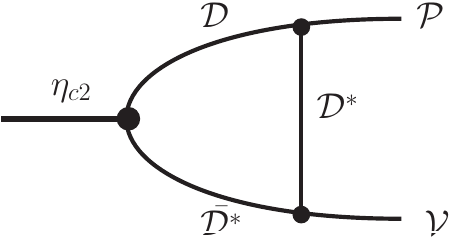}&
		\includegraphics[width=2.8cm]{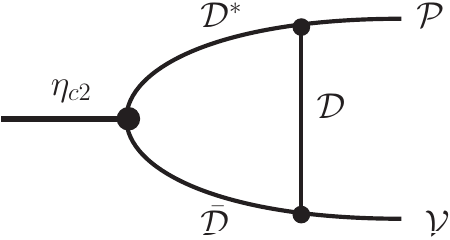}&
		\includegraphics[width=2.8cm]{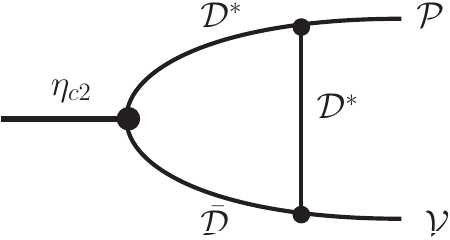}\\
		\\
		$(K_1)$ & $(K_2)$ & $(K_3)$ \\
		\\
		\includegraphics[width=2.8cm]{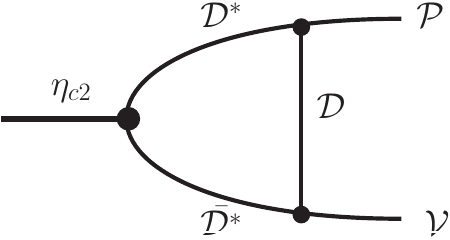}&
		\includegraphics[width=2.8cm]{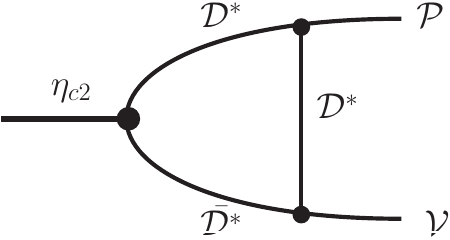}&
	 \\
		\\
		$(K_4)$ & $(K_5)$ &   \\
	\end{tabular}
	\caption{Diagrams contributing to $\eta_{c2} \to \mathcal{PV}$ at the hadron level .}\label{Fig:Tri42}
\end{figure}

Similar to the processes $\psi(1^3D_J)\to \mathcal{P}\mathcal{V}$, the meson loop contributions to the processes $\eta_{c2}\to \mathcal{P}\mathcal{V}$ are listed in Fig.~\ref{Fig:Tri42}. The amplitudes of the process $\eta_{c2}\to \mathcal{P}\mathcal{V}$, corresponding to Fig.~\ref{Fig:Tri42}-$(K_1)$-$(K_5)$, reads, 

\begin{eqnarray}
	\mathcal{M}_{K_1}&=&i^3 \int\frac{d^4 q}{(2\pi)^4}
	\Big[ig_{\eta_{c2} \mathcal{D}^{*} \mathcal{D}} \varepsilon^{\mu\alpha}(p)\left(p_{1\mu}- p_{2\mu}\right)\Big]\nonumber\\
	&&\left[g_{ \mathcal{D}^{*} \mathcal{D}P} p_{3\chi }\right]\left[g_{\mathcal{D}^{*} \mathcal{D}^{*}V}g_{\tau \phi} \epsilon^{\tau}\left(p_{4}\right)\left(q^{\xi}-p^{\xi}_{2}\right)\right.\nonumber\\
	&&\left.-4f_{\mathcal{D}^{*} \mathcal{D}^{*}V} \epsilon^{\tau}\left(p_{4}\right)\left(p_4^{\xi}g_{\tau \phi}-p^{\phi}_{4}g_{\tau \xi}\right)\right]\nonumber\\
	&& \times\frac{1}{p^2_1-m^2_1} \frac{-g^{\alpha\xi}+p_2^{\alpha} p_2^{\xi} /m^2_2}{p^2_2-m^2_2}\nonumber\\
	&&\times\frac{-g^{\chi\phi}+q^{\chi} q^{\phi} /m^2_q}{q^2-m^2_q}\mathcal{F}^2(m_q^2,q^2),\nonumber\\ 
	\mathcal{M}_{K_2}&=&i^3 \int\frac{d^4 q}{(2\pi)^4}
	\Big[ig_{\eta_{c2} \mathcal{D}^{*} \mathcal{D}} \varepsilon^{\mu\alpha}(p)\left(p_{1\mu}- p_{2\mu}\right)\Big]\nonumber\\
	&&\left[-g_{ \mathcal{D}^{*} \mathcal{D}P} p_{3\chi }\right]\left[g_{ \mathcal{D}\mathcal{D}V} \epsilon^{\tau}\left(p_{4}\right) \left(p_{2\tau}-q_{\tau}\right)\right]\nonumber\\
	&& \times\frac{-g^{\chi\rho}+p_1^{\alpha} p_1^{\chi} /m^1_2}{p^2_1-m^2_1}\frac{1}{p^2_2-m^2_2}\nonumber\\
	&& \times\frac{1}{q^2-m^2_q}\mathcal{F}^2(m_q^2,q^2),\nonumber 
\end{eqnarray}
\begin{eqnarray}	
	\mathcal{M}_{K_3}&=&i^3 \int\frac{d^4 q}{(2\pi)^4}
	\Big[ig_{\eta_{c2} \mathcal{D}^{*} \mathcal{D}} \varepsilon^{\mu\alpha}(p)\left(p_{1\mu}- p_{2\mu}\right)\Big] \nonumber\\
	&&\left[\frac{1}{2} g_{ \mathcal{D}^{*} \mathcal{D}^{*}P } \varepsilon_{\chi\lambda\sigma\rho}  p_{3}^{\lambda} \left(p^{\sigma}_{1}+q^{\sigma}\right) \right]\nonumber\\
	&&\left[2f_{\mathcal{D}^*\mathcal{D}V}\varepsilon_{\kappa\tau\phi\xi}p_4^\kappa\epsilon^\tau(p_4)\big(p_2^\phi-q^\phi\big)\right]\nonumber\\
	&&\times\frac{-g^{\alpha\rho}+p_1^{\alpha} p_1^{\rho} /m^1_2}{p^2_1-m^2_1}\frac{1}{p^2_2-m^2_2} \nonumber\\
	&&\times\frac{-g^{\chi\phi}+q^{\chi} q^{\phi} /m^2_q}{q^2-m^2_q}\mathcal{F}^2(m_q^2,q^2),\nonumber\\
	\mathcal{M}_{K_4}&=&i^3  \int\frac{d^4 q}{(2\pi)^4}
	\Big[ig_{\eta_{c2} \mathcal{D}^{*} \mathcal{D}^{*}} \varepsilon_{\alpha\beta\nu\gamma}p^{\gamma} \epsilon^{\mu\nu}(p)(p_{2\mu}- p_{1\mu})\Big] \nonumber\\
	&&\left[-g_{ \mathcal{D}^{*} \mathcal{D}P} p_{3\chi }\right]\left[2f_{\mathcal{D}^*\mathcal{D}V}\varepsilon_{\kappa\tau\phi\xi}p_4^\kappa\epsilon^\tau(p_4)\big(q^\phi-p2^\phi\big)\right]\nonumber\\
	&&\times\frac{-g^{\alpha\chi}+p_1^{\alpha} p_1^{\chi} /m^2_1}{p^2_1-m^2_1} \frac{-g^{\beta\xi}+p_2^{\beta} p_2^{\xi} /m^2_2}{p^2_2-m^2_2}\nonumber\\
	&&\times\frac{1}{q^2-m^2_q}\mathcal{F}^2(m_q^2,q^2),\nonumber
\end{eqnarray}
\begin{eqnarray}
	\mathcal{M}_{K_5}&=&i^3 \int\frac{d^4 q}{(2\pi)^4}
	\Big[ig_{\eta_{c2} \mathcal{D}^{*} \mathcal{D}^{*}} \varepsilon_{\alpha\beta\nu\gamma}p^{\gamma} \epsilon^{\mu\nu}(p)(p_{2\mu}- p_{1\mu})\Big] \nonumber\\
	&&\left[\frac{1}{2} g_{ \mathcal{D}^{*} \mathcal{D}^{*}P } \varepsilon_{\chi\lambda\sigma\rho}  p_{3}^{\lambda} \left(p^{\sigma}_{1}+q^{\sigma}\right) \right]\nonumber\\
	&&\left[g_{\mathcal{D}^{*} \mathcal{D}^{*}V}g_{\tau \phi} \epsilon^{\tau}\left(p_{4}\right)\left(q^{\xi}-p^{\xi}_{2}\right)\right.\nonumber\\
	&&\left.-4f_{\mathcal{D}^{*} \mathcal{D}^{*}V} \epsilon^{\tau}\left(p_{4}\right)\left(p_4^{\xi}g_{\tau \phi}-p^{\phi}_{4}g_{\tau \xi}\right)\right]\nonumber\\
	&&\times\frac{-g^{\alpha\rho}+p_1^{\alpha} p_1^{\rho} /m^2_1}{p^2_1-m^2_1} \frac{-g^{\beta\xi}+p_2^{\beta} p_2^{\xi} /m^2_2}{p^2_2-m^2_2}\nonumber\\
	&&\times\frac{-g^{\chi\phi}+q^{\chi} q^{\phi} /m^2_q}{q^2-m^2_q}\mathcal{F}^2(m_q^2,q^2).\nonumber
\end{eqnarray}

Then, the total amplitude for $\eta_{c2} \to  \mathcal{P}\mathcal{V}$  reads,
\begin{eqnarray}
	\mathcal{M}_{\eta_{c2}\to \mathcal{P}\mathcal{V}} =\sum_{j=1}^{5}\mathcal{M}_{K_j}.
\end{eqnarray}

\bibliography{psi1D.bib}
\bibliographystyle{unsrt}
\end{document}